\input harvmac
\let\includefigures=\iftrue
\let\useblackboard==\iftrue
\newfam\black

\includefigures
\message{If you do not have epsf.tex (to include figures),}
\message{change the option at the top of the tex file.}
\input epsf
\def\figin{\epsfcheck\figin}\def\figins{\epsfcheck\figins}
\def\epsfcheck{\ifx\epsfbox\UnDeFiNeD
\message{(NO epsf.tex, FIGURES WILL BE IGNORED)}
\gdef\figin##1{\vskip2in}\gdef\figins##1{\hskip.5in}
\else\message{(FIGURES WILL BE INCLUDED)}%
\gdef\figin##1{##1}\gdef\figins##1{##1}\fi}
\def\DefWarn#1{}
\def\figinsert{\goodbreak\midinsert}
\def\ifig#1#2#3{\DefWarn#1\xdef#1{fig.~\the\figno}
\writedef{#1\leftbracket fig.\noexpand~\the\figno}%
\figinsert\figin{\centerline{#3}}\medskip\centerline{\vbox{
\baselineskip12pt\advance\hsize by -1truein
\noindent\footnotefont{\bf Fig.~\the\figno:} #2}}
\endinsert\global\advance\figno by1}
\else
\def\ifig#1#2#3{\xdef#1{fig.~\the\figno}
\writedef{#1\leftbracket fig.\noexpand~\the\figno}%
\global\advance\figno by1} \fi

\def\id{{1 \kern-.28em {\rm l}}}

\def\K3{{\bf K3}}
\def\journal#1&#2(#3){\unskip, \sl #1\ \bf #2 \rm(19#3) }
\def\andjournal#1&#2(#3){\sl #1~\bf #2 \rm (19#3) }

\def\bar{\overline}
\def\hat{\widehat}
\def\ie{{\it i.e.}}
\def\eg{{\it e.g.}}

\def\tilde{\widetilde}

\def\frac#1#2{{#1\over#2}}

\def\inbar{\,\vrule height1.5ex width.4pt depth0pt}
\def\IC{\relax\hbox{$\inbar\kern-.3em{\rm C}$}}
\def\IR{\relax{\rm I\kern-.18em R}}
\def\IP{\relax{\rm I\kern-.18em P}}

%
%

\def\ap#1#2#3{Ann. Phys. {\bf #1} (#2) #3}

\catcode`\@=11
\def\slash#1{\mathord{\mathpalette\c@ncel{#1}}}
\overfullrule=0pt

\def\underrel#1\over#2{\mathrel{\mathop{\kern\z@#1}\limits_{#2}}}

\catcode`\@=12


%

\def\tr{{\rm tr}}

\def \sinh{{\rm sinh}}
\def \cosh{{\rm cosh}}


\def\ie{{\it i.e.}}
\def\eg{{\it e.g.}}


\lref\MAsratT{
M.~Asrat,
``Moving holographic boundaries,''
Nucl. Phys. B {\bf 1008}, 116699 (2024)
doi:10.1016/j.nuclphysb.2024.116699
[arXiv:2305.15744 [hep-th]].
}

\lref\Giveon{
A.~Giveon, N.~Itzhaki and D.~Kutasov,
``$T{\bar T} $ and LST,''
JHEP {\bf 07}, 122 (2017)
doi:10.1007/JHEP07(2017)122
[arXiv:1701.05576 [hep-th]].
}

\lref\Deser{
S.~Deser and R.~Jackiw,
``Three-Dimensional Cosmological Gravity: Dynamics of Constant Curvature,''
Annals Phys. {\bf 153}, 405-416 (1984)
doi:10.1016/0003-4916(84)90025-3
}

\lref\DeserT{
S.~Deser, R.~Jackiw and G.~'t Hooft,
``Three-Dimensional Einstein Gravity: Dynamics of Flat Space,''
Annals Phys. {\bf 152}, 220 (1984)
doi:10.1016/0003-4916(84)90085-X
}

\lref\Banadoswn{
M.~Banados, C.~Teitelboim and J.~Zanelli,
``The Black hole in three-dimensional space-time,''
Phys. Rev. Lett. {\bf 69}, 1849-1851 (1992)
doi:10.1103/PhysRevLett.69.1849
[arXiv:hep-th/9204099 [hep-th]].
}

\lref\Banadosgq{
M.~Banados, M.~Henneaux, C.~Teitelboim and J.~Zanelli,
``Geometry of the (2+1) black hole,''
Phys. Rev. D {\bf 48}, 1506-1525 (1993)
[erratum: Phys. Rev. D {\bf 88}, 069902 (2013)]
doi:10.1103/PhysRevD.48.1506
[arXiv:gr-qc/9302012 [gr-qc]].
}

\lref\Horowitz{
G.~T.~Horowitz and D.~L.~Welch,
``Exact three-dimensional black holes in string theory,''
Phys. Rev. Lett. {\bf 71}, 328-331 (1993)
doi:10.1103/PhysRevLett.71.328
[arXiv:hep-th/9302126 [hep-th]].
}

\lref\Horne{
J.~H.~Horne and G.~T.~Horowitz,
``Exact black string solutions in three-dimensions,''
Nucl. Phys. B {\bf 368}, 444-462 (1992)
doi:10.1016/0550-3213(92)90536-K
[arXiv:hep-th/9108001 [hep-th]].
}

\lref\HorneS{
J.~H.~Horne, G.~T.~Horowitz and A.~R.~Steif,
``An Equivalence between momentum and charge in string theory,''
Phys. Rev. Lett. {\bf 68}, 568-571 (1992)
doi:10.1103/PhysRevLett.68.568
[arXiv:hep-th/9110065 [hep-th]].
}

\lref\Garfinkle{
D.~Garfinkle,
``Black string traveling waves,''
Phys. Rev. D {\bf 46}, 4286-4288 (1992)
doi:10.1103/PhysRevD.46.4286
[arXiv:gr-qc/9209002 [gr-qc]].
}

\lref\Sfetsos{
K.~Sfetsos,
``Conformally exact results for SL(2,R) x SO(1,1)(d-2) / SO(1,1) coset models,''
Nucl. Phys. B {\bf 389}, 424-444 (1993)
doi:10.1016/0550-3213(93)90327-L
[arXiv:hep-th/9206048 [hep-th]].
}

\lref\Bars{
I.~Bars and K.~Sfetsos,
``Exact effective action and space-time geometry n gauged WZW models,''
Phys. Rev. D {\bf 48}, 844-852 (1993)
doi:10.1103/PhysRevD.48.844
[arXiv:hep-th/9301047 [hep-th]].
}

\lref\CarlipT{
S.~Carlip and C.~Teitelboim,
``Aspects of black hole quantum mechanics and thermodynamics in (2+1)-dimensions,''
Phys. Rev. D {\bf 51}, 622-631 (1995)
doi:10.1103/PhysRevD.51.622
[arXiv:gr-qc/9405070 [gr-qc]].
}

\lref\Carlip{
S.~Carlip,
``The (2+1)-Dimensional black hole,''
Class. Quant. Grav. {\bf 12}, 2853-2880 (1995)
doi:10.1088/0264-9381/12/12/005
[arXiv:gr-qc/9506079 [gr-qc]].
}

\lref\Coussaert{
O.~Coussaert and M.~Henneaux,
``Supersymmetry of the (2+1) black holes,''
Phys. Rev. Lett. {\bf 72}, 183-186 (1994)
doi:10.1103/PhysRevLett.72.183
[arXiv:hep-th/9310194 [hep-th]].
}

\lref\Witten{
E.~Witten,
``On string theory and black holes,''
Phys. Rev. D {\bf 44}, 314-324 (1991)
doi:10.1103/PhysRevD.44.314
}

\lref\Mandal{
G.~Mandal, A.~M.~Sengupta and S.~R.~Wadia,
``Classical solutions of two-dimensional string theory,''
Mod. Phys. Lett. A {\bf 6}, 1685-1692 (1991)
doi:10.1142/S0217732391001822
}

\lref\Chandrasekhar{
Chandrasekhar Subrahmanyan and Hartle J.~ B.~, 
``On crossing the Cauchy horizon of a Reissner-Nordstrom black-hole,''
Proc. R. Soc. Lond. A {\bf 384}, 301-315 (1982)
doi:10.1098/rspa.1982.0160
}

\lref\Callan{
C.~G.~Callan, Jr., E.~J.~Martinec, M.~J.~Perry and D.~Friedan,
``Strings in Background Fields,''
Nucl. Phys. B {\bf 262}, 593-609 (1985)
doi:10.1016/0550-3213(85)90506-1
}

\lref\Chan{
K.~C.~K.~Chan and R.~B.~Mann,
``Spinning black holes in (2+1)-dimensional string and dilaton gravity,''
Phys. Lett. B {\bf 371}, 199-205 (1996)
doi:10.1016/0370-2693(95)01609-0
[arXiv:gr-qc/9510069 [gr-qc]].
}

\lref\ChanK{
K.~C.~K.~Chan,
``Comment on the calculation of the angular momentum for the (anti)selfdual charged spinning BTZ black hole,''
Phys. Lett. B {\bf 373}, 296-298 (1996)
doi:10.1016/0370-2693(96)00145-1
[arXiv:gr-qc/9509032 [gr-qc]].
}

\lref\BrownY{
J.~D.~Brown and J.~W.~York, Jr.,
``Quasilocal energy and conserved charges derived from the gravitational action,''
Phys. Rev. D {\bf 47}, 1407-1419 (1993)
doi:10.1103/PhysRevD.47.1407
[arXiv:gr-qc/9209012 [gr-qc]].
}

\lref\Brown{
J.~D.~Brown, J.~Creighton and R.~B.~Mann,
``Temperature, energy and heat capacity of asymptotically anti-de Sitter black holes,''
Phys. Rev. D {\bf 50}, 6394-6403 (1994)
doi:10.1103/PhysRevD.50.6394
[arXiv:gr-qc/9405007 [gr-qc]].
}

\lref\Kamata{
M.~Kamata and T.~Koikawa,
``2+1 dimensional charged black hole with (anti-)self dual Maxwell fields,''
Phys. Lett. B {\bf 353}, 87-92 (1995)
doi:10.1103/PhysRevD.50.6394
[arXiv:gr-qc/9405007 [gr-qc]].
}

\lref\Antoniadis{
I.~Antoniadis, C.~Bachas, J.~R.~Ellis and D.~V.~Nanopoulos,
``Cosmological String Theories and Discrete Inflation,''
Phys. Lett. B {\bf 211}, 393-399 (1988)
doi:10.1016/0370-2693(88)91882-5
}

\lref\Abbott{
L.~F.~Abbott and S.~Deser,
``Stability of Gravity with a Cosmological Constant,''
Nucl. Phys. B {\bf 195}, 76-96 (1982)
doi:10.1016/0550-3213(82)90049-9
}

\lref\HawkingA{
S.~W.~Hawking and G.~T.~Horowitz,
``The Gravitational Hamiltonian, action, entropy and surface terms,''
Class. Quant. Grav. {\bf 13}, 1487-1498 (1996)
doi:10.1088/0264-9381/13/6/017
[arXiv:gr-qc/9501014 [gr-qc]].
}

\lref\KomarA{
A.~Komar,
``Covariant conservation laws in general relativity,''
Phys. Rev. {\bf 113}, 934-936 (1959)
doi:10.1103/PhysRev.113.934
}

\lref\YorkBC{
J.~York,
``Boundary terms in the action principles of general relativity,''
Found. Phys. {\bf 16}, 249-257 (1986)
doi:10.1007/BF01889475
}

\lref\Gibbons{
G.~W.~Gibbons and S.~W.~Hawking,
``Action Integrals and Partition Functions in Quantum Gravity,''
Phys. Rev. D {\bf 15}, 2752-2756 (1977)
doi:10.1103/PhysRevD.15.2752
}

\lref\Hayward{
G.~Hayward,
``Gravitational action for space-times with nonsmooth boundaries,''
Phys. Rev. D {\bf 47}, 3275-3280 (1993)
doi:10.1103/PhysRevD.47.3275
}

\lref\Arnowitt{
R.~L.~Arnowitt, S.~Deser and C.~W.~Misner,
``The Dynamics of general relativity,''
Gen. Rel. Grav. {\bf 40}, 1997-2027 (2008)
doi:10.1007/s10714-008-0661-1
[arXiv:gr-qc/0405109 [gr-qc]].
}

\lref\Kodama{
H.~Kodama,
``Conserved Energy Flux for the Spherically Symmetric System and the Back Reaction Problem in the Black Hole Evaporation,''
Prog. Theor. Phys. {\bf 63}, 1217 (1980)
doi:10.1143/PTP.63.1217
}

\lref\Kinoshita{
S.~Kinoshita,
``Extension of Kodama vector and quasilocal quantities in three-dimensional axisymmetric spacetimes,''
Phys. Rev. D {\bf 103}, no.12, 124042 (2021)
doi:10.1103/PhysRevD.103.124042
[arXiv:2103.07408 [gr-qc]].
}

\lref\Bardeen{
J.~M.~Bardeen, W.~H.~Press and S.~A.~Teukolsky,
``Rotating black holes: Locally nonrotating frames, energy extraction, and scalar synchrotron radiation,''
Astrophys. J. {\bf 178}, 347 (1972)
doi:10.1086/151796
}

\lref\BardeenP{
J.~M.~Bardeen,
``A Variational Principle for Rotating Stars in General Relativity,''
Astrophys. J. {\bf 162}, 71 (1970)
doi:10.1086/150635
}

\lref\Eghbali{
A.~Eghbali, L.~Mehran-nia and A.~Rezaei-Aghdam,
``BTZ black hole from Poisson-Lie T-dualizable sigma models with spectators,''
Phys. Lett. B {\bf 772}, 791-799 (2017)
doi:10.1016/j.physletb.2017.07.044
[arXiv:1705.00458 [hep-th]].
}

\lref\Edery{
A.~Edery,
``Non-singular vortices with positive mass in 2+1 dimensional Einstein gravity with AdS$_3$ and Minkowski background,''
JHEP {\bf 01}, 166 (2021)
doi:10.1007/JHEP01(2021)166
[arXiv:2004.09295 [hep-th]].
}

\lref\Regge{
T.~Regge and C.~Teitelboim,
``Role of Surface Integrals in the Hamiltonian Formulation of General Relativity,''
Annals Phys. {\bf 88}, 286 (1974)
doi:10.1016/0003-4916(74)90404-7
}

\lref\Keski{
E.~Keski-Vakkuri,
``Bulk and boundary dynamics in BTZ black holes,''
Phys. Rev. D {\bf 59}, 104001 (1999)
doi:10.1103/PhysRevD.59.104001
[arXiv:hep-th/9808037 [hep-th]].
}

\lref\Balasubramanian{
V.~Balasubramanian, P.~Kraus and A.~E.~Lawrence,
``Bulk versus boundary dynamics in anti-de Sitter space-time,''
Phys. Rev. D {\bf 59}, 046003 (1999)
doi:10.1103/PhysRevD.59.046003
[arXiv:hep-th/9805171 [hep-th]].
}

\lref\Font{
A.~Font, L.~E.~Ibanez, D.~Lust and F.~Quevedo,
``Strong - weak coupling duality and nonperturbative effects in string theory,''
Phys. Lett. B {\bf 249}, 35-43 (1990)
doi:10.1016/0370-2693(90)90523-9
}

\lref\Buscher{
T.~H.~Buscher,
``A Symmetry of the String Background Field Equations,''
Phys. Lett. B {\bf 194}, 59-62 (1987)
doi:10.1016/0370-2693(87)90769-6
}

\lref\BuscherB{
T.~H.~Buscher,
``Path Integral Derivation of Quantum Duality in Nonlinear Sigma Models,''
Phys. Lett. B {\bf 201}, 466-472 (1988)
doi:10.1016/0370-2693(88)90602-8
}

\lref\Rocek{
M.~Rocek and E.~P.~Verlinde,
``Duality, quotients, and currents,''
Nucl. Phys. B {\bf 373}, 630-646 (1992)
doi:10.1016/0550-3213(92)90269-H
[arXiv:hep-th/9110053 [hep-th]].
}

\lref\BerkovitsB{
N.~Berkovits and J.~Maldacena,
``Fermionic T-Duality, Dual Superconformal Symmetry, and the Amplitude/Wilson Loop Connection,''
JHEP {\bf 09}, 062 (2008)
doi:10.1088/1126-6708/2008/09/062
[arXiv:0807.3196 [hep-th]].
}

\lref\FDavidMM{
F.~David,
``Conformal Field Theories Coupled to 2D Gravity in the Conformal Gauge,''
Mod. Phys. Lett. A {\bf 3}, 1651 (1988)
doi:10.1142/S0217732388001975
}

\lref\Alvarez{
E.~Alvarez, L.~Alvarez-Gaume, J.~L.~F.~Barbon and Y.~Lozano,
``Some global aspects of duality in string theory,''
Nucl. Phys. B {\bf 415}, 71-100 (1994)
doi:10.1016/0550-3213(94)90067-1
[arXiv:hep-th/9309039 [hep-th]].
}

\lref\FradkinT{
E.~S.~Fradkin and A.~A.~Tseytlin,
``Quantum String Theory Effective Action,''
Nucl. Phys. B {\bf 261}, 1-27 (1985)
[erratum: Nucl. Phys. B {\bf 269}, 745-745 (1986)]
doi:10.1016/0550-3213(85)90559-0
}

\lref\Knizhnik{
V.~G.~Knizhnik and A.~B.~Zamolodchikov,
``Current Algebra and Wess-Zumino Model in Two-Dimensions,''
Nucl. Phys. B {\bf 247}, 83-103 (1984)
doi:10.1016/0550-3213(84)90374-2
}

\lref\TseytlinPR{
A.~A.~Tseytlin,
``Effective action of gauged WZW model and exact string solutions,''
Nucl. Phys. B {\bf 399}, 601-622 (1993)
doi:10.1016/0550-3213(93)90511-M
[arXiv:hep-th/9301015 [hep-th]].
}

\lref\Chandrasekaran{
V.~Chandrasekaran, E.~E.~Flanagan, I.~Shehzad and A.~J.~Speranza,
``Brown-York charges at null boundaries,''
JHEP {\bf 01}, 029 (2022)
doi:10.1007/JHEP01(2022)029
[arXiv:2109.11567 [hep-th]].
}

\lref\Graves{
J.~C.~Graves and D.~R.~Brill,
``Oscillatory Character of Reissner-Nordstrom Metric for an Ideal Charged Wormhole,''
Phys. Rev. {\bf 120}, 1507-1513 (1960)
doi:10.1103/PhysRev.120.1507
}

\lref\Lunin{
O.~Lunin and P.~Shah,
``Double Field Theory and $\alpha'$ corrections: explicit examples,''
[arXiv:2408.04833 [hep-th]].
}

\lref\Kaloper{
N.~Kaloper and K.~A.~Meissner,
``Duality beyond the first loop,''
Phys. Rev. D {\bf 56}, 7940-7953 (1997)
doi:10.1103/PhysRevD.56.7940
[arXiv:hep-th/9705193 [hep-th]].
}

\lref\HorowitzZZZ{
G.~T.~Horowitz and A.~Strominger,
``Black strings and P-branes,''
Nucl. Phys. B {\bf 360}, 197-209 (1991)
doi:10.1016/0550-3213(91)90440-9
}

\Title{
} {\vbox{
{\vbox{
\centerline{Kalb-Ramond field, black holes and } }}
\smallskip
\centerline{ black strings in (2 + 1)D} }}

\bigskip
\centerline{\it Meseret Asrat}
\smallskip
\centerline{International Center for Theoretical Sciences}
\centerline{Tata Institute of Fundamental Research
} \centerline{Bengaluru, KA 560089, India}

\smallskip

\vglue .3cm

\bigskip

\let\includefigures=\iftrue
\bigskip
\noindent

New rotating dilaton black hole and black string solutions in three spacetime dimensions are obtained. The background spacetime interpolates between Anti-de Sitter and a (an asymptotically) flat spacetime. The new black strings are characterized by their masses, angular momenta and axion charges. The uncharged black string solutions have only a single horizon. Enclosed inside their horizons, they contain a curvature singularity. Thus, they are also black hole solutions of Einstein-scalar gravity. On the other hand, the charged black string solutions have two horizons. They may or may not contain a curvature singularity depending on the ratio of their inner and outer horizons radii. When they contain a singularity, the singularity is either at or enclosed inside their inner horizons. We also obtain, in addition to the uncharged black strings, other new black hole solutions of Einstein-scalar gravity by applying duality transformation on the new charged black strings. The new black holes are described by their masses and angular momenta. We show that their angular momenta do not depend on their horizons radii. We also find that the masses of a class of the black holes can be made negative by adjusting, using the gauge ambiguity, the asymptotic value of the (independent component of the) Kalb-Ramond field in the dual black strings. We also discuss black hole and black string solutions with a curvature singularity at or beyond their outer or event horizons. Novel black hole solutions with a ring curvature singularity in between their inner and outer horizons are also presented.

\bigskip

\Date{09/24}

\newsec{Introduction}

In this paper, we obtain new black hole and black string solutions in the string background obtained in \MAsratT. The string background has the product topology ${\cal A}_3 \times S^3 \times {\cal X}_4$, where $S^3$ denotes a unit three-sphere and ${\cal X}_4$ is a compact four dimensional manifold. Examples of ${\cal X}_4$ include the Calabi-Yau manifolds $T^4$ and $K3$. The spacetime ${\cal A}_3$ interpolates between $AdS_3$ in the infrared and a (an asymptotically) linear dilaton spacetime $\IR\times S^1\times \IR$ in the ultraviolet. The world-sheet conformal field theory on ${\cal A}_3$ is an exact marginal current bilinear deformation of the world-sheet theory on $AdS_3$. In the boundary theory the deformation is equivalent to a deformation by an irrelevant operator of (left and right) scaling dimensions $(2, 2)$, see \refs{\MAsratT,\ \Giveon} and references therein. The resulting deformed theory is thus non-conformal and dual to string theory on ${\cal A}_3$ (times the internal product manifold ${\cal M}_7 := S^3 \times {\cal X}_4$).

In the boundary theory the deformation coupling $\hat\gamma$ has mass dimension $-2$. The coupling can have either signs. In the case it is negative the deformation washes the holographic boundary away to infinity. On the other hand, in the case it is positive, in general it moves the boundary into the bulk. In Anti-de Sitter ($AdS$) spacetime moving the boundary into the bulk generates at onset a curvature singularity. For more and thorough discussions see \MAsratT. Therefore, we will make in our following discussions a distinction between positive and negative coupling. Also since products of conformal field theories (CFTs) are still CFTs, we will simply ignore the (world-sheet) CFT on the internal manifold ${\cal M}_7$. Therefore, ${\cal M}_7$ is unaffected throughout our discussion.

The string background ${\cal A}_3$ contains the Kalb-Ramond field. The field has only one independent component. At asymptotic infinity in general the component is a non-zero constant. We can freely set the constant to any arbitrary value using the gauge ambiguity. As we will show later in the paper, the usual gauge ambiguity in the definition of the Kalb-Ramond field or the constant value that its (independent) component takes at the boundary, \ie\ at radial infinity, has a non-trivial consequence. In particular, it allows for black hole solutions (of Einstein-scalar gravity) with negative ADM (Arnowitt-Deser-Misner) masses.\foot{In a situation in which the metric (in the Einstein frame) is not asymptotically flat and/or the cosmological constant is not zero the usual ADM mass formula does not directly apply. In such a situation, the adapted or analogous ADM mass \Abbott, however, is still given by the usual ADM expression except that now the reference metric is not flat and the lapse function is not unity. It equals the asymptotic value at radial infinity of the Brown-York (quasi-local) mass \HawkingA. In this paper, we follow the Brown-York approach \BrownY\ and simply call the Brown-York mass at radial infinity the physical or ADM mass. See Appendix C for a review on the Brown-York energy and mass.}

In three spacetime dimensions the Hodge dual of the Kalb-Ramond field is a (pseudo-)scalar which is identified as an axion field \Antoniadis. Its value at infinity gives an additional parameter to describe the solutions.\foot{T-duality relates it to angular momentum.} The solutions we find are asymptotically flat (in the string frame) and they are exact solutions of classical string theory.

The new black string solutions are parametrized by their masses, angular momenta and axion charges. The axion charge is the value of the (pseudo-)scalar at the boundary or spatial infinity. In general it is determined by the electric flux in the world-volume of the black string. We show that the axion charge is given only in terms of the deformation coupling. In particular it does not depend on the horizons (radii). The uncharged solutions, \ie, those black string solutions with zero axion charge, are also black hole solutions of Einstein-scalar gravity. We also obtain other new black hole solutions of Einstein-scalar gravity by applying duality transformation on the new charged black string solutions.\foot{See Appendix B for a review of the duality transformation.} We show that their angular momenta are given by the dual black strings axion charges and thus do not depend on the horizons. We also show that the masses of a class of the black hole solutions can be made negative by changing (in the dual black strings) the value of the (component of the) Kalb-Ramond field at radial infinity.

The uncharged black strings have only a single horizon. They contain a curvature singularity with the structure of a ring. In the case the deformation coupling is negative the singularity is enclosed inside the event horizon.\foot{This particular class of solutions, \ie, the class of solutions characterized by zero axion charge and negative coupling (of $-1$), is already known in the literature \Chan. It is obtained by directly solving the Einstein field equations with a minimally coupled scalar field. However, here we simply obtain it using duality and coordinate transformations from global $AdS_3$.  We will discuss it in more detail later in the paper.} Otherwise, \ie, for positive coupling, the singularity is beyond the event horizon. The charged black strings, on the other hand, possess two horizons. We show that depending on the ratio of the two horizons radii the solutions may or may not contain a curvature singularity. The singularity has the structure of a ring. In the case the deformation coupling is negative the singularity is at or enclosed inside the inner horizon.\foot{The inner (event) horizon is also referred to as Cauchy horizon.} Otherwise, it is at or beyond the outer horizon. The new black hole solutions that we obtain by applying duality transformation have only a single horizon. They contain a curvature singularity. The singularity is enclosed inside the event horizon. To end, the black hole solutions that we will discuss, including those characterized by positive coupling, have only a single horizon and a ring singularity. The event horizon divides the spacetime into two regions. The interior region contains the singularity and the exterior region contains the rest of the spacetime. This will be our definition of interior and exterior. In the case the coupling is positive the ring singularity cannot be contracted to the (coordinate) origin without crossing the horizon(s).

The paper is organized as follows. In section two we review the rotating BTZ black hole \refs{\Banadoswn,\ \Banadosgq}. In section three we obtain the new (dilaton) black hole and black string solutions. In section four we further discuss the new (dilaton) black hole solutions. They are (black hole) solutions of a minimally coupled Einstein-scalar gravity theory. We in particular compute their ADM masses and angular momenta. In section five we summarize the main results and discuss future research directions. We also give, in section five, novel black hole solutions with a ring curvature singularity in between their inner and outer horizons.

In Appendix A we review a parameterization of $AdS_3$ that in particular is better suited in relation to BTZ black hole. In Appendix B we review T-duality, also known as target-space duality. We use the T-duality transformation in section three. In Appendix C we review the Brown-York quasi-local energy and mass. In Appendix D we review the Kruskal approach. We use it to remove the coordinate singularities of the solutions we study in the paper and to obtain their Penrose diagrams.  

\newsec{BTZ black hole}

Unlike in higher dimensional gravity, in three dimensional gravity, the Riemann curvature tensor is determined completely by the Einstein tensor or equivalently the energy momentum tensor of the matter content.\foot{The Weyl curvature tensor is also identically zero. Thus, there are no physical excitations or propagating gravitons.} Therefore, in the absence of matter the only solution is (locally) flat spacetime. However, in the case the cosmological constant is non-zero and negative, it has black hole solution \refs{\Banadoswn,\ \Banadosgq}. The solution is referred to as BTZ black hole. It is a solution of Einstein equations in three spacetime dimensions with negative cosmological constant and without matter or fields.

The BTZ black hole is related to global $AdS_3$ (\ie, the universal cover of the $SU(1, 1) \cong SL(2, \IR)$ group manifold,) by a global coordinate transformation. $AdS_3$ (which here always denotes the infinitely-sheeted universal covering space of $SU(1, 1) \cong SL(2, \IR)$) is described in global coordinates by the metric
\eqn\za{
ds^2 = l^2(d\theta^2 - \cosh^2\theta d\varphi^2 + \sinh^2\theta d\psi^2),
} 
where $l$ is the radius of curvature and it is related to the (negative) cosmological constant $\Lambda$. $\psi$ is an angle variable and it has period $2\pi$. The time coordinate $\varphi$ takes its values in $\IR$ and thus it is not compact. The radial coordinate $\theta$ takes its values in $[0, \infty)$. We next discuss the coordinate transformation and obtain the BTZ black hole from $AdS_3$.

We first make the coordinate redefinition
\eqn\zb{
r = l\sinh\theta.
}
The metric \za\ becomes 
\eqn\zc{
ds^2 = \frac{dr^2}{\frac{r^2}{l^2} + 1} - \left(r^2 + l^2\right)d\varphi^2 + r^2 d\psi^2,
}
We next make the coordinate transformation \refs{\Banadosgq\CarlipT-\Carlip}\foot{The complex number $i$ is related to a change of basis or coordinate patch, see Appendix A and \refs{\Balasubramanian,\ \Keski}.}
\eqn\zd{\matrix{
\varphi  = i\bar\varphi, & \bar\varphi = -\left(\frac{\rho_-}{l^2}t - \frac{\rho_+}{l} \omega\right),\cr
\psi = i\bar\psi, & \bar\psi =  -\left(\frac{\rho_+}{l^2}t - \frac{\rho_-}{l} \omega\right),
}
}
where $t, \omega \in \IR$, and extend the domain of $r$ to all the real numbers, \ie, $r = {\bar r} \in \IR$, see A.5. We define, since we are in particular interested in the region exterior to the black hole horizons, the new coordinate $\rho$ as
\eqn\zdx{
 {\bar r}^2  = l^2\left(\frac{\rho^2 - \rho_+^2}{\rho_+^2 - \rho_-^2}\right),
}
where $\rho \geq \rho_+ > \rho_-$. In our convention the coordinates $\rho$ and $t$ have mass dimension $-1$. Using both the $GL(2)$ transformation \zd\ and the change of variable \zdx\ in \zc\ we get the metric
\eqn\ze{\eqalign{
ds^2 & =\frac{l^2\rho^2d\rho^2}{(\rho^2 - \rho_+^2)(\rho^2 - \rho_-^2)} - \frac{(\rho^2 - \rho_+^2)(\rho^2 - \rho_-^2)}{l^2\rho^2}dt^2 + \rho^2\left(d\omega - \frac{\rho_+\rho_-}{l\rho^2}dt\right)^2,\cr
& = \frac{l^2\rho^2d\rho^2}{(\rho^2 - \rho_+^2)(\rho^2 - \rho_-^2)} - \frac{(\rho^2 - \rho_+^2 - \rho^2_-)}{l^2}dt^2 - 2\frac{\rho_+\rho_-}{l} dt d\omega + \rho^2 d\omega^2.
}
}
The metric is invariant under $\rho \to -\rho$. The BTZ black hole metric is obtained after identifying $\omega$ periodically. In the metric \ze\ we now identify $\omega$ with period $2\pi$, \ie, $\omega \sim \omega + 2\pi$. This in turn implies $(\bar\varphi, \bar\psi) \sim (\bar\varphi + 2\pi \rho_+/l, \bar\psi + 2\pi \rho_-/l)$. Now $\rho$ is a radial coordinate, \ie\ $\rho \geq 0$. Therefore, BTZ black hole can be viewed as a quotient space of $AdS_3$. Thus, locally it is isometric to $AdS_3$. Furthermore, the scalar curvature is constant. It is given by\foot{In $AdS_d$ the scalar curvature and cosmological constant are given by
 \eqn\adsdr{
    R = -\frac{d(d - 1)}{l^2}, \quad \Lambda = -\frac{(d - 1)(d - 2)}{2l^2}.
}
}
\eqn\btzr{
    R = -\frac{6}{l^2} = 6\cdot \Lambda, \quad \Lambda = -\frac{1}{l^2},
}
where $\Lambda$ is the cosmological constant. Asymptotically, the BTZ black hole approaches (without identifications) $AdS_3$ spacetime.

The solution has two horizons. It has an event horizon at $\rho_+$ and an inner horizon at $\rho_-$. There is no curvature singularity inside the inner horizon. The spacetime region,
\eqn\bm{\rho^2_+ < \rho^2 < \rho_+^2 + \rho_-^2,
} 
in which the Killing vector field $(\partial_t)^\mu = (1, 0, 0) := \delta^\mu_t$ is space-like defines an ergosphere.\foot{The analytically extended BTZ metric is obtained by replacing $\rho$ with $i\rho$ or simply by setting $\rho^2 = r \in \IR$. It contains closed time-like curves and thus usually it is not considered. $r = 0$ is a causal singularity.}

The BTZ black hole metric \ze\ satisfies the vacuum Einstein equations in $(2 + 1)$ spacetime dimensions,
\eqn\einst{
    G_{ab} := R_{ab} - \frac{1}{2}Rg_{ab} = -\Lambda g_{ab} = \frac{1}{l^2}g_{ab}.
}
In an asymptotically flat spacetime the total mass (which equals the total energy since the lapse is one) is unambiguously defined. It is given by the ADM mass. The mass in $AdS$ and more generally in non-flat metrics is given by the usual ADM expression except that now the reference metric is not flat and the lapse is not one \refs{\Abbott, \HawkingA}. It equals the Brown-York quasi-local mass \BrownY\ at radial infinity. The Brown-York mass is measured by a static observer. The angular momentum is given similarly by the Brown-York quasi-local angular momentum at radial infinity. It agrees with the Komar angular momentum \KomarA, here adapted for $(2 + 1)$ dimensions.

The angular momentum and (analogous) ADM mass of the BTZ solution are given by\foot{In our convention, the mass $M$ is dimensionless. We set Newton's gravitational constant $G_N$ to $1/8$ (in units of length). The angular momentum $J$ has mass dimension $-1$. For BTZ black hole the usual Komar mass formula adapted for $(2 + 1)$ dimensions gives a divergent answer.}
\eqn\admmnj{
    J = \frac{2\rho_+\rho_-}{l}, \quad M = \frac{\rho_+^2 + \rho_-^2}{l^2}.
}
In terms of $M$ and $J$ the BTZ (black hole) metric is
\eqn\zex{
ds^2  =\frac{l^2\rho^2d\rho^2}{(\rho^4 - l^2 M\rho^2  + {l^2J^2\over 4})} - \frac{(\rho^4 - l^2 M\rho^2  + {l^2J^2\over 4})}{l^2\rho^2}dt^2 + \rho^2\left(d\omega - \frac{J}{2\rho^2}dt\right)^2.
}
Thus, it is described by two parameters, its mass $M$ and angular momentum $J$. We will obtain $M$ and $J$ later in the paper following the Brown-York approach. In order for the solution to describe a black hole or for a horizon to exist we assume
\eqn\bl{M >0,\quad |J| \leq Ml.
}
It is with these conditions that the BTZ metric contains black hole solutions.

We note that $(M, J) = (-1, 0)$ is the $AdS_3$ metric \zc. It is separated from the massless BTZ solution, \ie, $(M, J) = (0, 0)$, or the continuous black hole spectrum, by a mass unit. The solutions with $-1 < M < 0, J = 0$ have naked conical singularity at the origin $\rho = 0$. For small $\rho$ the metric is
\eqn\conical{ds^2 = d{\tilde \rho}^2 - d\tau^2 -M{\tilde \rho}^2 d\omega^2, \quad \rho = \sqrt{-M}{\tilde \rho}, \quad  \tau = \sqrt{-M}t.
}
Thus, it describes point particle sources with negative cosmological constant \refs{\Deser, \ \DeserT}.\foot{See \Edery\ for numerical solutions of static non-singular vortices in $AdS_3$ that are not BTZ black holes.} The zero mass black hole is identified in $(1, 1)$ $AdS$ supergravity \Coussaert\ as the Ramond sector ground state. Global $AdS_3$ is identified as the Neveu-Schwarz sector ground state.

In the next section we obtain a new family of rotating black hole and black string solutions by applying to ${\cal A}_3$, since in the infrared the spacetime ${\cal A}_3$ is well described by $AdS_3$, the same $GL(2)$ transformation. We explain this in the next section.

\newsec{Rotating black hole and black string solutions}
\subsec{New black string solutions}
We now consider the string background ${\cal A}_3$ obtained in \MAsratT. ${\cal A}_3$ contains the metric $g_{\mu\nu}$, the Kalb-Ramond field $B_{\mu\nu}$ and the dilaton field $\Phi$. The string metric $g_{\mu\nu}$ is given by
\eqn\aaa{\eqalign{
ds^2 & = g_{ab}dx^adx^b,\cr
& = l^2(d\theta^2 - e^{2\phi}\cosh^2\theta d\varphi^2 + e^{2\phi}\sinh^2\theta d\psi^2),
}
}
where
\eqn\bbb{
e^{-2\phi} = 1 + \gamma^2 - 2\gamma\cosh (2\theta),
}
and $\gamma$ is a dimensionless parameter. It is related to the deformation coupling $\hat\gamma$ in the dual boundary field theory by $\hat\gamma = \gamma R^2$ (up to a constant factor) here $R$ is the conformal radius of the boundary cylinder on which the undeformed CFT is defined.\foot{The conformal boundary of $AdS_3$ (\ie, the universal cover) is a time-like cylinder $\IR \times S^1$. As usual in AdS, we take $R = 1$ in string units.} The two form $B$ is given by
\eqn\ccc{
B = B_{01}d\varphi\wedge d\psi, \quad B_{01} = -\frac{1}{2}l^2e^{2\phi}(\gamma - \cosh(2\theta)).
}
The dilaton $\Phi$ is given by 
\eqn\ddd{
e^{2\Phi} = g_s^2 |e^{2\phi}| = {g_s^2\over \left|1 + \gamma^2 - 2\gamma \cosh(2\theta)\right|},
}
where $g_s$ is the string coupling at $\gamma = 0$. Note the absolute value sign is necessary in the case $\gamma$ is positive to ensure the dilaton is real everywhere. More on this momentarily. The Kalb-Ramond field strength $H$ is given by
\eqn\dddx{\eqalign{
H & = dB,\cr
& = l^2 (1 - \gamma^2)\sinh(2\theta)e^{4\phi} d\varphi\wedge d\psi\wedge d\theta.
}
}
The time coordinate $\varphi$ takes its value in $\IR$. The radial coordinates $\theta$ takes its value (in the case $\gamma \leq 0$) in $\IR^+$, \ie\ $0\leq \theta < \infty$. The angular variable $\psi$ is periodic. It has period $2\pi(1 - \gamma)$ to ensure that there is no conical singularity at $\theta = 0$. The deformation coupling $\gamma$ takes the values in the range
\eqn\zf{-1 \leq \gamma \leq 1.
}
Other values of $\gamma$ can be mapped onto the interval \zf\ by rescaling the coordinates and the string coupling at $\gamma = 0$, \ie, $g_s$. The transformation rules under $\gamma \to 1/\gamma$ are $\varphi \to \varphi/\gamma$, $\psi \to -\psi/\gamma$ and $g_s^2 \to g_s^2/\gamma^2$ (see Appendix E in \MAsratT). The metric, the three form antisymmetric tensor and dilaton are invariant by these transformations.

In a moment, we replace $\theta$ using a change of variable by a function of a new coordinate $\rho$. When $\gamma > 0$, the range of the radial coordinate $\rho$ is restricted so that the argument of the absolute value \ddd\ is always positive. Therefore, in this paper one may simply ignore the absolute value sign. For a detailed discussion on the range of $\theta$ when $\gamma > 0$, see \MAsratT.

$AdS_3$ (with Kalb-Ramond two form) is an exact solution to (classical) string theory, \ie, there are no $\alpha' = l_s^2$ corrections to $AdS_3$. $l_s$ is the string length. The solution ${\cal A}_3$ is related to $AdS_3$ by (abelian) T-dualities and coordinate shift (see \MAsratT\ for details), and thus it is exact in $\alpha'$. However, it is perturbative in $g_s$ \Font. In general, an abelian T-duality maps an exact solution of string theory into an exact one (see, \eg, \Lunin). However, in the case the solution is an effective one, abelian T-duality can generate $\alpha'$ corrections \Kaloper. Later in the section we give previously known $\alpha'$ exact (black string) solutions as special cases of our results.

On the world-sheet the deformation corresponds to an exact marginal deformation (see, \eg, \MAsratT). More on this in a future work.

In the low energy (super)gravity theory the metric, three form flux and dilaton are governed by the action
\eqn\sgracTT{S = {1\over 2\kappa^2}\int d^3 x\sqrt{-g}e^{-2\tilde\Phi}\left[R + 4\left(\nabla \tilde\Phi \right)^2 - {1\over 12}H^2 - 2\Lambda\right],
}
where $x^\mu = (x^0, x^1, x^2) = (\varphi, \psi, \theta)$ and
\eqn\gsgnNN{e^{\tilde\Phi} = {e^{\Phi}\over g_s}, \quad {1\over 2\kappa^2} = {1\over 2\kappa_0^2 g_s^2} = {1\over 16\pi G_N}, \quad \Lambda = -{2\over l^2},
}
where $G_N$ is the gravitational coupling in three spacetime dimensions. The equations of motion that follow from the low energy (super)gravity action are
\eqn\eqmotionTT{\eqalign{R_{\mu\nu} + 2\nabla_\mu\nabla_\nu\tilde\Phi - {1\over 4}H_{\mu\rho\sigma}H_{\nu}^{\rho\sigma} = 0,\cr
\nabla^{\rho}\left(e^{-2\tilde\Phi} H_{\mu\nu\rho}\right) = 0,\cr
R + 4\nabla^2\tilde\Phi - 4\left(\nabla\tilde\Phi\right)^2 - {1\over 12}H^2 - 2\Lambda = 0,
}
}
where $R = g^{\mu\nu}R_{\mu\nu}$ is the (string frame) Ricci scalar.

For $\gamma = 0$, ${\cal A}_3$ is $AdS_3$. Assuming $\gamma$ is non-zero, for small $\theta$, \ie\ $\sinh\theta \ll 1$, the metric \aaa\ describes $AdS_3$ and for large $\theta$, \ie\ $\sinh\theta \gg 1$ the metric is asymptotically flat. In what follows we simply apply the same $GL(2)$ transformation to obtain black hole and black string solutions in ${\cal A}_3$.

We now first make the coordinate redefinition
\eqn\eee{
r = l\sinh\theta.
}
Using this the fields $g_{\mu\nu}$,  $B_{\mu\nu}$ and $\Phi$ become
\eqn\fff{\eqalign{
ds^2 & = \frac{dr^2}{\frac{r^2}{l^2} + 1} - e^{2\phi}\left(r^2 + l^2\right)d\varphi^2 + e^{2\phi} r^2 d\psi^2,\cr
B & = B_{01}d\varphi\wedge d\psi, \quad B_{01} = -\frac{1}{2}l^2e^{2\phi}\left(\gamma - 1 - 2\frac{r^2}{l^2}\right),\cr
e^{2\Phi} & = g_s^2 |e^{2\phi}|,
}
}
where
\eqn\ggg{
e^{-2\phi} = (1 - \gamma)^2 - 4\gamma\frac{r^2}{l^2}.
}

We next make the coordinate transformation,
\eqn\jjj{\eqalign{
\varphi & = -i\left(\frac{\rho_-}{l^2}t - \frac{\rho_+}{l} x\right),\cr
\psi & =  -i\left(\frac{\rho_+}{l^2}t - \frac{\rho_-}{l} x\right),
}
}
and the change of variable
\eqn\jjjz{
r^2 = l^2\left(\frac{\rho^2 - \rho_+^2}{\rho_+^2 - \rho_-^2}\right).
}

The $GL(2)$ transformation \jjj\ and the coordinate redefinition \jjjz\ give the black string solution described by the metric $g_{\mu\nu}$, Kalb-Ramond field $B_{\mu\nu}$ and dilaton field $\Phi$ which are given by
\eqn\mmm{\eqalign{
ds^2 & = g_{\mu\nu}dx^\mu dx^\nu\cr
& = \frac{l^2\rho^2d\rho^2}{(\rho^2 - \rho_+^2)(\rho^2 - \rho_-^2)} - e^{2\phi}\frac{(\rho^2 - \rho_+^2)(\rho^2 - \rho_-^2)}{l^2\rho^2}dt^2 + e^{2\phi}\rho^2\left(dx - \frac{\rho_+\rho_-}{l\rho^2}dt\right)^2,\cr
& = \frac{l^2\rho^2d\rho^2}{(\rho^2 - \rho_+^2)(\rho^2 - \rho_-^2)} - e^{2\phi}\frac{(\rho^2 - \rho_+^2 - \rho_-^2)}{l^2}dt^2 - 2e^{2\phi}\frac{\rho_+\rho_-}{l}dxdt + e^{2\phi}\rho^2dx^2,\cr
B & = B_{tx}dt\wedge dx, \quad B_{tx} = \frac{1}{2}e^{2\phi}\left(\frac{\rho^2_+ - \rho^2_-}{l}\right)\left[\gamma - 1 - 2\left(\frac{\rho^2 - \rho^2_+}{\rho^2_+ - \rho^2_-}\right)\right],\cr
e^{2\Phi} & = g_s^2 |e^{2\phi}|.
}
}
where
\eqn\ooo{\eqalign{
e^{-2\phi} & = (1 - \gamma)^2 - 4\gamma\left(\frac{\rho^2 - \rho^2_+}{\rho^2_+ - \rho^2_-}\right),\cr
& = \frac{\rho_+^2 (1 + \gamma)^2 - \rho_-^2(1 - \gamma)^2 - 4\gamma \rho^2}{\rho_+^2 - \rho_-^2}.
}
}
We will shortly explain why we refer to the solution as black string. The time coordinate $t$ takes its value in $\IR$. We will in general assume $x$ is an angle coordinate. $\rho$ is a radial coordinate. We will in general also assume, unless stated otherwise, that $\rho_- < \rho_+$. The periodicity of $x$ is chosen so that there is no conical singularity at the origin. We note that at $\rho = \infty$ the $B$ field, \ie, $B_{tx}$, is constant. We can adjust the constant using the gauge freedom. This has, as we will show in the next section, a non-trivial consequence. In particular, the constant enters into physical observables. 

The three form flux is
\eqn\fluxfintif{\eqalign{H & = dB,\cr
& = H_{t x \rho} dt \wedge dx \wedge d\rho,
}
}
where
\eqn\bq{\eqalign{
H_{t x \rho} & = \frac{\partial B_{t x}}{\partial \rho}, \cr
& = -\frac{2(1 - \gamma^2)}{l}\cdot \frac{(\rho_+^2 - \rho_-^2)^2\rho}{\left[\rho_+^2(1 + \gamma)^2 - \rho_-^2(1 - \gamma)^2 - 4\gamma \rho^2\right]^2},\cr
& = -\frac{2(1 - \gamma^2)}{l}\cdot e^{4\phi}\cdot \rho = \frac{2(1 - \gamma^2)}{l}\cdot \left(\frac{g}{\rho}\right),
}
}
where $g$ is the determinant of the string metric $g_{\mu\nu}$. 

The black string solutions carry the axion charge (per unit length) $Q$ given by the value of the pseudo-scalar or axion field $e^{-2\Phi} {\star H}$ at the boundary or spatial infinity. It follows from the (super)gravity equations of motion, \ie, $\nabla^\rho(e^{-2\Phi}H_{\mu\nu\rho}) = 0$, that the axion is a constant and therefore $Q$ is simply given by 
\eqn\bp{Q = e^{-2\Phi} {\star H} = 2\left(\frac{1 - \gamma^2}{lg_s^2}\right),
}
where the Hodge dual $\star$ is taken with respect to the three dimensional metric \mmm. We have used $\varepsilon_{tx\rho} = +1$ where $\varepsilon_{tx\rho}$ is the totally antisymmetric tensor. We have also assumed, for a reason that will be clear in a moment, $-1 \leq \gamma \leq 0$.\foot{In general, for $\gamma \leq 0$, the charge is $Q = e^{-2\Phi} {\star H} = 2\left|\frac{1 - \gamma^2}{lg_s^2}\right|$. The solutions (labelled by the triplet $(\rho_-, \rho_+; \gamma)$ are invariant under $\gamma \to 1/\gamma$. Thus, $Q$ is invariant under $\gamma \to 1/\gamma$. This is in particular self-evident because of the absolute value sign.} We note that the charge $Q$ depends only on $\gamma$. In particular, it does not depend on $\rho_+$ and $\rho_-$.

We can always put the solution \mmm, for $\gamma \neq 0$, into the form 
\eqn\newFFF{\eqalign{
ds^2 & = L^{-2}(\rho)d\rho^2 - N^2(\rho)d{\hat t}^2 + {(\rho^2_+ - \rho^2_-)\over 4|\gamma|}\left(d{\hat x} + Zd{\hat t}\right)^2,\cr
B & = B_{{\hat t}{\hat x}}d{\hat t}\wedge d{\hat x}, \quad B_{{\hat t}{\hat x}} = \left(\frac{\rho_+^2 - \rho_-^2}{2l}\right)\cdot \left[\frac{\rho_+^2(1 + \gamma) + \rho_-^2(1 - \gamma) - 2\rho^2}{\rho_+^2(1 + \gamma)^2 - \rho_-^2(1 - \gamma)^2 - 4\gamma\rho^2}\right],\cr
e^{2\Phi} & = g_s^2 \left|{\rho_+^2 - \rho_-^2\over \rho_+^2(1 + \gamma)^2 - \rho_-^2(1 - \gamma)^2 - 4\gamma\rho^2}\right|,
}
}
by unwrapping $x$ and applying a coordinate transformation. The metric components, \ie, $L$, $N$ and $Z$, are 
\eqn\newFFFMC{\eqalign{
L^2 & = {(\rho^2 - \rho_-^2)(\rho^2 - \rho_+^2)\over l^2 \rho^2},\cr
N^2 & = {4|\gamma| (\rho_+^2 - \rho_-^2) \over l^2}\cdot {(\rho^2 - \rho_-^2)(\rho^2 - \rho_+^2)\over [\rho_+^2(1 + \gamma)^2 - \rho_-^2(1 - \gamma)^2 - 4\gamma\rho^2]^2},\cr
Z &= \left(\gamma\over |\gamma|\right)\cdot \left(1 - \gamma^2\over l\right) \cdot \left[{\rho_+^2 - \rho_-^2 \over \rho_+^2(1 + \gamma)^2 - \rho_-^2(1 - \gamma)^2 - 4\gamma\rho^2}\right].
}
}
For $\gamma < 0$, the coordinates are related by the Lorentz boost $t = {\hat t}\cosh(\xi) + l{\hat x}\sinh(\xi), lx = l{\hat x}\cosh(\xi) + {\hat t}\sinh(\xi)$, where $\xi = (1/2)\cdot \ln \{|\gamma|\cdot \left[(\rho_+ + \rho_-)/(\rho_+ - \rho_-)\right]\}$. Alternatively, we can take $\xi = (1/2)\cdot \ln \{(1/|\gamma|)\cdot \left[(\rho_+ + \rho_-)/(\rho_+ - \rho_-)\right]\}$. The latter choice only changes $Z \to -Z$. Note the $B$ field is invariant under the boost. Thus, also the axion charge $Q$ is invariant. For $\gamma > 0$, the coordinates are instead related by the (general) coordinate transformation $lx = -{\hat t}\cosh(\xi) + l{\hat x}\sinh(\xi), t = l{\hat x}\cosh(\xi) - {\hat t}\sinh(\xi)$, where $\xi$ is as given above.\foot{Note the (general) coordinate transformation, for positive $\gamma$, is not a Lorentz (boost) transformation. It is a combination of a Lorentz boost and a spatial-temporal rotation (of $\pi/2$). We obtain the boost transformation by making a spacetime rotation, \ie, the replacements ${\hat x} \to {\hat t}/l, {\hat t} \to -l{\hat x}$.} Also for $\gamma > 0$, the latter choice for $\xi$ only changes $Z \to -Z$. The $B$ field, also for $\gamma > 0$, is invariant under the (general) coordinate transformation.

The black hole solution \newFFF\ extends (or winds around) and, for non-constant $Z$, also rotates along the (compact) ${\hat x}$ direction. For this reason, therefore, we refer to the solution \mmm\ as rotating black string. When $Z$ is a constant, we simply redefine ${\hat x}$ as ${\hat x} \to {\hat x} - Z{\hat t}$ to effectively set $Z$ to zero. Therefore, for $Z = 0$ (or, more generally, for constant $Z$), the black hole is a trivial bundle over (the spatial dimension) ${\hat x}$ and, it is commonly referred to as simply black string. Note that a black string does not always necessarily contain a black hole factor. That is, the fiber may not describe or solve a black hole solution. However, in a boosted and rotated frame or in a different coordinate system it could describe a black hole. A simple example is the solution at $\gamma = -1$. At $\gamma = -1$, we note that $Z = 0$. The solution has the product form ${\cal M}_2 \times \IR$. The factor ${\cal M}_2$ does not describe a black hole in two dimensions. However, as we momentarily show, we can define a new coordinate system in which the solution contains a two dimensional black hole factor.  Also, a product spacetime of a black hole solution and a line $\IR$ or a circle $S^1$ does not always necessarily solve the string theory equations of motion and thus, it does not always necessarily describe or lead to a black string solution \HorowitzZZZ.

The solution \newFFF\ is related to the black string solution \Horne. However, the relation is not direct. We explain this now. To this end, it is convenient to introduce (assuming $\gamma \neq 0$) the new coordinate ${\hat \rho}$,
\eqn\shifTTT{4|\gamma|{\hat\rho}^2 = -4\gamma\rho^2  + \left[\rho_+^2(1 + \gamma)^2 - \rho_-^2(1 - \gamma)^2\right].
}
Note that by shifting (and/or analytically continuing) $\rho^2$ we are effectively fixing $|\gamma|$ to be equal to $(\rho_+ - \rho_-)/(\rho_+ + \rho_-)$. This will become more clear later in the section. In terms of the new coordinates $({\hat t}, {\hat x}, {\hat \rho})$ the solution \newFFF\ takes the form
\eqn\nnewFFF{\eqalign{
ds^2 & = { l^2 {\hat \rho}^2 d{\hat \rho}^2 \over ({\hat \rho}^2 - {\hat \rho}_-^2)({\hat \rho}^2 - {\hat \rho}_+^2)} - {{\hat\rho}^2_+ \over \left(1 + |\gamma|\right)^2}\left[({\hat\rho}^2 - {\hat\rho}^2_+)({\hat\rho}^2 - {\hat\rho}^2_-) \over  l^2 {\hat\rho}^4\right]d{\hat t}^2 +  {{\hat\rho}^2_+ \over \left(1 + |\gamma|\right)^2}\left[d{\hat x} + \left({\gamma\over|\gamma|}\right) {{\hat\rho}_-{\hat\rho}_+\over l {\hat\rho}^2}d{\hat t}\right]^2,\cr
B & = B_{{\hat t}{\hat x}}d{\hat t}\wedge d{\hat x}, \quad B_{{\hat t}{\hat x}} = \left(\frac{|\gamma|}{\gamma}\right) \cdot {1 \over \left(1 + |\gamma|\right)^2} \cdot {{\hat\rho}^2_+ \over l} \cdot \left[1 - {{\hat \rho}_+{\hat \rho}_-\over {\hat \rho}^2}\right],\cr
e^{2\Phi} & = {g_s^2\over \left(1 + |\gamma|\right)^2} \left({{\hat \rho}_+^2\over {\hat \rho}^2}\right),
}
}
where ${\hat \rho}_+$ and ${\hat \rho}_-$ are given in terms of $\rho_+$, $\rho_-$ and $\gamma$,
\eqn\ewDEFF{{\hat\rho}^2_+ = {(1 + |\gamma|)^2\over 4|\gamma|}(\rho_+^2 - \rho_-^2), \quad {\hat\rho}^2_- = {(1 - |\gamma|)^2\over 4|\gamma|}(\rho_+^2 - \rho_-^2).
}
Note we have effectively avoided the need for the absolute value sign in \ddd\ when $\gamma > 0$ by shifting $\rho$ \shifTTT, or equivalently, by restricting $\rho$ to the region $\rho < \rho_+$. We have also shrunk/moved the curvature singularity to the origin. Moreover, we can eliminate the explicit dependence of \nnewFFF\ on $\gamma$ by redefining $g_s$ and the two coordinates ${\hat t}$ and ${\hat x}$. Therefore, the dependence on $\gamma$ is implicit. To obtain this implicit dependence we take a closer look at \ewDEFF. The relations \ewDEFF\ imply the constraint\foot{Upon an appropriate change of coordinates, the solution \nnewFFF\ becomes \eqn\nnewCCO{\eqalign{
ds^2 & = {l^2\over 4}\left[{dr^2\over (r - r_-)(r - r_+)} - {(r - r_-)(r - r_+)\over r^2}d\tau^2 + \left(d\chi + {\gamma\over |\gamma|}{\sqrt{r_-r_+}\over r}d\tau\right)^2\right],\cr
B & = B_{\tau\chi}d\tau\wedge d\chi, \quad B_{\tau\chi} = {l^2\over 4}\left[1 - {\sqrt{r_-r_+}\over r}\right],\cr
\Phi & = {1\over 2}\ln\left(r_+\over r\right) + \iota, \quad \iota = \ln\left({g_s\over 1 + |\gamma|}\right), \quad {r_-\over r_+} = \left({1 - |\gamma|\over 1 + |\gamma|}\right)^2.
}
}
The scalar curvature is $R = {2\over l^2r^2}\left[2(r_+ + r_-)r - 7r_+r_-\right]$. Thus, it is singular at the origin $r = 0$.
}
\eqn\ewDEFFCON{{{\hat\rho}_-\over {\hat\rho}_+} = {1 - |\gamma|\over 1 + |\gamma|}, \quad |\gamma| = {{\hat\rho}_+ - {\hat\rho}_-\over {\hat\rho}_+ + {\hat\rho}_-}.
}
Thus, the ratio ${\hat\rho}_-/{\hat\rho}_+$ is fixed by $\gamma$. In particular, at $\gamma = -1$ we have ${\hat\rho}_- = 0$. Therefore, in the new coordinates, the solution at $\gamma = -1$ has the product form ${\cal M}_2 \times \IR$ where ${\cal M}_2$ is a black hole \refs{\Witten, \ \Mandal}. Later, we will explicitly show that \nnewFFF\ is related indirectly to the non-rotating black string solution \Horne\ by a further Lorentz boost.\foot{In \nnewFFF\ we could treat ${\hat\rho}_\pm$ more generally as independent parameters and simply ignore their dependence on $\gamma$ \ewDEFF\ or the constraint \ewDEFFCON. However, this is a separate case.} To the best of my knowledge the black string solution \mmm\ has not been reported elsewhere in the literature.

Note at $\gamma = 0$ the metric describes BTZ black hole and it allows a non-zero Kalb-Ramond field strength $H$ \Horowitz.\foot{Therefore, one may as well prefer to refer to the solutions (labelled by the triplet $(\rho_-, \rho_+; \gamma)$) \mmm\ as deformed black holes.} The BTZ black hole solution \ze\ now with $\Phi = 0$ and $B = B_{tx} dt\wedge dx$ where $B_{tx} = -\left[\rho^2 - (\rho_+^2 + \rho_-^2)/2\right]/l$, \ie, the solution \mmm\ at $\gamma = 0$, is dual to the charged black string solution \Horowitz\foot{In the appropriate gauge and with the appropriate coordinate changes.}, see also \refs{\Horne, \ \HorneS, \ \Garfinkle, \ \Sfetsos, \ \Bars} and \Eghbali. We next consider different cases of the black string solution described by \mmm. We begin with the case where $\gamma$ is negative.\foot{The analytically extended solution, which is obtained by simply replacing $\rho$ by $i\tilde\rho$ or $\rho^2$ by $r \in \IR$, in general contains closed time-like curves and thus we will not consider it. In the Penrose diagrams below we also do not include the regions which contain closed time-like curves.}

\subsec{Negative coupling}

In this case we consider $-1 < \gamma < 0$. The scalar curvature is finite provided the ratio 
\eqn\bbbb{\frac{\rho_-}{\rho_+} < \mu := \frac{1 + \gamma}{1 - \gamma}.
}
Thus, in the case $\rho_-/\rho_+ < \mu$ we have rotating black string solutions with inner and outer horizons but no curvature singularity. The solutions are exact solutions of classical string theory since they are related to ${\cal A}_3$ primarily by coordinate transformations. $\rho_-$ denotes the inner horizon and $\rho_+$ denotes the outer horizon. The period of the angle $x$ is chosen to avoid a conical singularity at the origin. We identify 
\eqn\ooox{x \sim x + 2\pi \sqrt{\frac{\rho_+^2(1 + \gamma)^2 - \rho_-^2(1 - \gamma)^2}{\rho^2_+ - \rho^2_-}}.
}
The global structure of the solutions which satisfy the inequality \bbbb\ is described using Penrose diagram in Fig. 1. The solutions are axially symmetric. Therefore, we set the (shifted) angular coordinate (in each patch) to a particular value, see Appendix D for details. Thus, a point in the diagram represents a circle in spacetime, \ie\ the orbit of the axial symmetry. The diagram extends infinitely in both vertical directions.
\ifig\loc{The plot depicts the global structure of the black string solutions with $\rho_- < \mu \cdot \rho_+$, where $\mu = (1 + \gamma)/(1 - \gamma), -1 < \gamma < 0$. The shifted angular coordinate is fixed in each of the patches to a particular value. $\rho_\pm$ represent horizons. The solutions have no singularity. Null geodesics are represented by straight lines at $\pi/4$ radians. In region II $\rho$ is time-like.}
{\epsfxsize2.4in\epsfbox{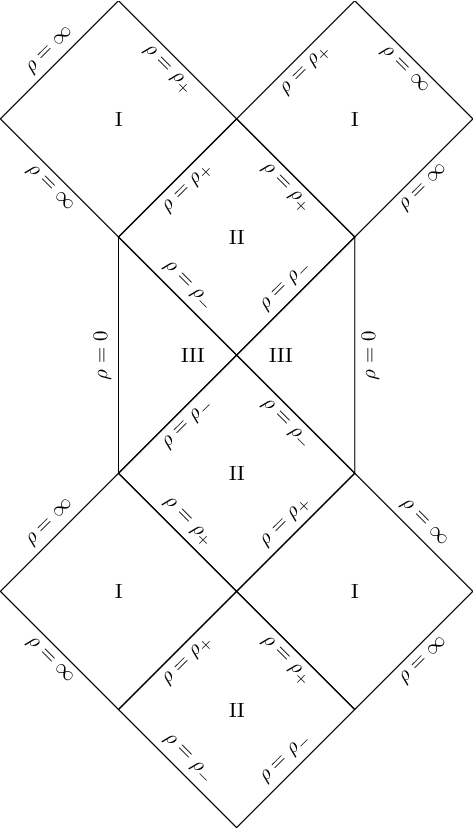}}
 
In the cases the condition \bbbb\ is not satisfied the scalar curvature is divergent at $\rho = \rho_0$ where
\eqn\cccc{\rho_0 = \frac{1}{2}\sqrt{\frac{\rho_+^2 (1 + \gamma)^2 - \rho_-^2(1 - \gamma)^2}{\gamma}}.
}
We note that at $\gamma = -1$ we have $\rho_0 = \rho_-$. Thus, at $\gamma = -1$ the inner horizon becomes singular and the black string solution has only one event horizon. Since $\rho_- < \rho_+$, the curvature singularity is enclosed inside the event horizon. In general, for negative values of $\gamma$ and $\rho_-/\rho_+ \geq \mu$, \ie, $-1 \leq \gamma \leq (\rho_- - \rho_+)/(\rho_- + \rho_+)$, we have $0 \leq \rho_0 \leq \rho_-$. $\rho_0 = 0$ when $\gamma = (\rho_- - \rho_+)/(\rho_- + \rho_+)$. Thus, the singularity is at or enclosed inside the inner horizon.

We now consider the special cases $\rho_0 = 0$ and $\rho_0 = \rho_-$. We first consider the case $\rho_0 = 0$. In this case we have
\eqn\spechorhowr{\gamma = {\rho_- - \rho_+\over \rho_- + \rho_+}, \quad \rho_+ > \rho_-.
}
The black string solution \mmm\ reduces to 
\eqn\spMwww{\eqalign{
ds^2 & = \frac{l^2\rho^2 d\rho^2}{(\rho^2 - \rho_+^2)(\rho^2 - \rho^2_-)} - {\rho_+^2\over (1 - \gamma)^2 }\frac{(\rho^2 - \rho_+^2)(\rho^2 - \rho^2_-)}{l^2\rho^4}dt^2 + \frac{\rho^2_+}{(1 - \gamma)^2}\left(dx - {\rho_+ \rho_-\over l \rho^2}dt\right)^2,\cr
B & = B_{tx}dt\wedge dx, \quad B_{tx} = -\frac{1}{(1 - \gamma)^2}\frac{\rho_+^2}{l}\left(1 - {\rho_+\rho_-\over \rho^2}\right), \quad Q = 2\left({1 - \gamma^2\over l g_s^2}\right),\cr
e^{2\Phi} & = g_s^2|e^{2\phi}| = {g_s^2\over (1 - \gamma)^2}\left(\rho_+^2\over \rho^2\right).
}
}
We first note that \spMwww\ is the same as \nnewFFF. The Ricci scalar is singular at the origin.\foot{Thus, here $x$ need not be periodic (with a definite period to avoid a conical singularity). Although we assumed $\rho_+ > \rho_-$, \ie\ $\gamma \neq 0$, to arrive at the solution, also here the limit $\rho_+ \to \rho_-$ is mathematically well-defined and gives a non-trivial solution. In this limit, after an appropriate change of coordinates, we obtain $ds^2 = (l^2/4)\left[dr^2/\left(r - r_0\right)^2 - d\tau^2\left(1 - (r_0/r)\right)^2 + \left(d\chi - (r_0/r)d\tau\right)^2\right]$, $B = B_{\tau\chi} d\tau\wedge d\chi$, $B_{\tau\chi} = -(l^2/4)\left[1 - (r_0/r)\right]$, $e^{2\Phi} = g_s^2r_0/r$, where $r_0 = \rho_+^2 = \rho_-^2$. The scalar curvature is $R = \left[2(4r - 7r_0)r_0\right]/(l^2r^2)$. Thus, $r = r_0$ is not a curvature singularity. We do not discuss this solution further here.} We also note that both the dilaton and Kalb-Ramond fields are singular at the origin. We will discuss this family of solutions further later in the section. We in particular show that they are related indirectly to the solutions discussed in \Horne. The global structure of the solutions for which $0 \leq \rho_0 < \rho_-$ is described using Penrose diagram in Fig. 2. The diagram is obtained in the same way as the example in Appendix D. In the diagram we set, in each patch, the (shifted) angular coordinate to a particular value since any other value is equivalent and related by axial symmetry. Therefore, a point in the diagram represents a circle in spacetime.
\ifig\loc{The plot depicts the global structure of the black string solutions with $\rho_- \geq \mu \cdot \rho_+$, where $\mu = (1 + \gamma)/(1 - \gamma), -1 < \gamma < 0$. $\rho_\pm $ represent horizons. The solutions have a singularity at $\rho = \rho_0$. $\rho_0$ is represented by the zigzag lines. In general $0 \leq \rho_0 < \rho_-$. The singularity is time-like. For $\rho_0 = 0$, the maximally extended solution has no closed time-like curves. Therefore, for $\rho_0 = 0$, the diagram can be extended behind $\rho = \rho_0 = 0$.}
{\epsfxsize2.4in\epsfbox{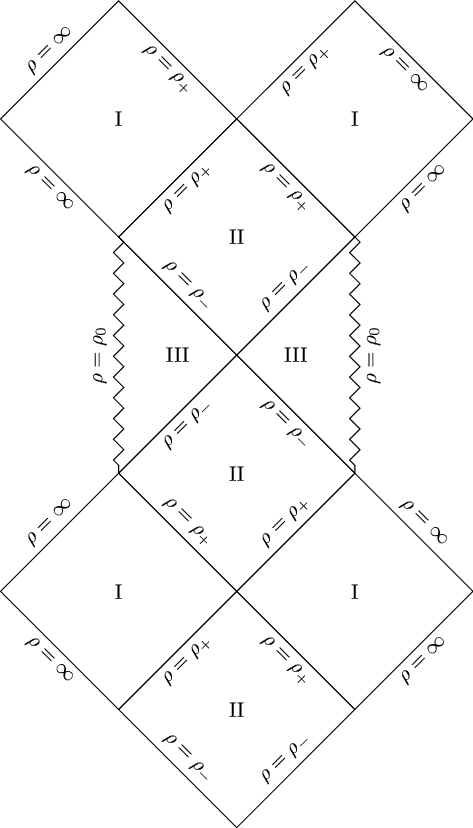}}

Now consider the case $\rho_0 = \rho_-$, \ie, $\gamma = -1$. The black string solution \mmm\ reduces at $\gamma = -1$ to
\eqn\ppp{\eqalign{
ds^2 & =\frac{l^2\rho^2d\rho^2}{(\rho^2 - \rho_+^2)(\rho^2 - \rho_-^2)} - \frac{1}{4}\frac{(\rho^2 - \rho_+^2)(\rho_+^2 - \rho_-^2)}{l^2\rho^2}dt^2 + \frac{1}{4}\frac{\rho^2(\rho_+^2 - \rho_-^2)}{\rho^2 - \rho_-^2}\left(dx - \frac{\rho_+\rho_-}{l\rho^2}dt\right)^2,\cr
B_{tx} & = -\frac{1}{4}\left(\frac{\rho_+^2 - \rho_-^2}{l}\right) = {\rm const.}, \quad Q = 0,\cr
e^{2\Phi} & = g_s^2|e^{2\phi}|, \quad e^{-2\phi} = 4\left(\frac{\rho^2 - \rho_-^2}{\rho_+^2 - \rho_-^2}\right).
}
}
The scalar curvature is given by
\eqn\sss{
R = \frac{4}{l^2}\left(\frac{\rho_+^2 - \rho_-^2}{\rho^2 - \rho_-^2}\right).
}
Note the $B$ field, \ie\ $B_{tx}$, is a non-zero constant. In general, using the gauge freedom we can set $B$ to zero. Thus, the solution describes a rotating dilaton black hole. This particular solution is already known in the literature \Chan. It is obtained by directly solving the Einstein field equations with a minimally coupled scalar field. The novelty here, however, is that we simply obtained it using duality and coordinate transformations from global $AdS_3$ \MAsratT. More on this black hole solution in the next section. The global structure of the solutions is described using Penrose diagram in Fig. 3. Each point represents a circle in spacetime.
\ifig\loc{The plot depicts the global structure of the black hole solutions with $\gamma = -1$.  $\rho_+$ represents the event horizon. The solutions have a space-like singularity at $\rho = \rho_-$. $\rho_-$ is represented by zigzag lines.}
{\epsfxsize2.8in\epsfbox{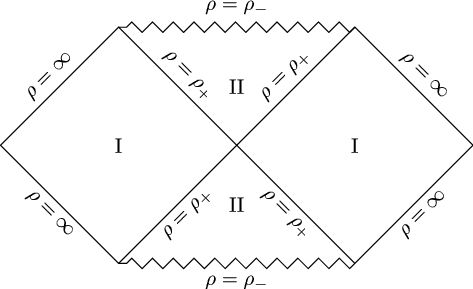}} 

We note that in the case $\gamma = -1$ and $\rho_- = 0$ the solution \mmm\ (or either \spMwww\ or \ppp\ with $\rho_- = 0$) simplifies to
\eqn\www{\eqalign{
ds^2 & = \frac{l^2d\rho^2}{(\rho^2 - \rho_+^2)} - \frac{\rho_+^2(\rho^2 - \rho_+^2)}{4l^2\rho^2}dt^2 + \frac{1}{4}\rho^2_+dx^2,\cr
B & = B_{tx}dt\wedge dx, \quad B_{tx} = -\frac{1}{4}\frac{\rho_+^2}{l} = {\rm const.}, \quad Q = 0,\cr
e^{2\Phi} & = {g_s^2\over 4}\left(\frac{\rho^2_+}{\rho^2}\right).
}
}
The curvature singularity is at the origin. This solution in the gauge in which $B = 0$ is dual to the non-rotating BTZ black hole with a non-vanishing $B$ \Horowitz.\foot{As we will show later in the paper, for a general gauge, it is also equivalent to a non-rotating black hole in a spacetime with non-vanishing angular velocity at radial infinity and no $B$ field.} It is also related indirectly to \Horne, see \HorneS. It is the product of the two dimensional black hole \refs{\Witten, \ \Mandal} with $S^1$. To see this define
\eqn\hhhh{
    \frac{\rho}{\rho_+} = \cosh\theta \geq 1.
}
The metric \www\ becomes
\eqn\iiii{
    ds^2 = l^2 (d\theta^2 - \tanh^2\theta d\tau^2 + d\psi^2),
}
where
\eqn\jjjj{
    \tau := \frac{\rho_+}{2l} t, \quad \psi := \frac{\rho_+}{2}x.
}
The dilaton is
\eqn\kkkk{
    e^{-2\Phi} = \frac{4}{g_s^2}\cosh^2\theta.
}
The scalar curvature is
\eqn\llll{
    R = \frac{4}{l^2}{\rm sech}^2\theta.
}
The analytic continuation $\tau \to i\tau$ and identification $\tau \sim \tau + 2\pi$ give a metric which describes a manifold with the topology of a semi-cigar times a circle.

\subsec{Positive coupling}

In the case $0 < \gamma \leq 1$ the scalar curvature is always divergent at $\rho = \rho_0$ \cccc, see \ooo. For $\gamma$ arbitrarily close to zero and positive, the metric describes BTZ black hole with a ring singularity at $\rho_0 = \infty$. Therefore, from the onset the deformation develops curvature singularity at the boundary \MAsratT. In general for positive $\gamma$, $\rho_0 \geq \rho_+$. Thus, the singularity is either at or beyond the outer horizon. Therefore, it cannot be continuously contracted or deformed to the origin without crossing the horizon(s). Since we are assuming that $x$ is compact, the solution has closed time-like curves behind the singularity in the interior region, \ie, $\rho > \rho_0$. The Penrose diagram for $\rho_0 > \rho_+$ is given in Fig. 4.
\ifig\loc{The plot depicts the global structure of the black string solutions with $0 < \gamma < 1$.  $\rho_\pm$ represent the horizons. $\rho = 0$ is a causal singularity. It is similar to the causal singularity in BTZ black hole. The solutions have a time-like singularity at $\rho = \rho_0$. $\rho_0$ is represented by zigzag lines. For arbitrarily small positive $\gamma$, the singularity is at arbitrarily large $\rho = \rho_0$.}
{\epsfxsize1.2in\epsfbox{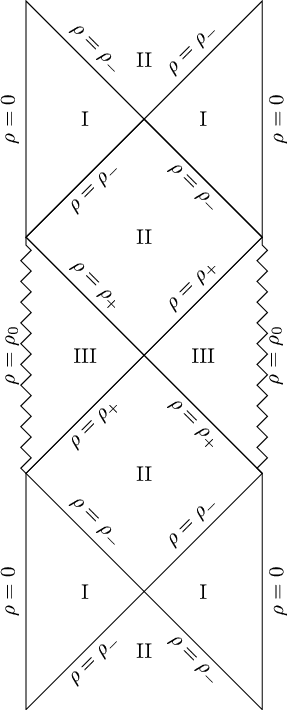}} 

We note that at $\gamma = 1$ we have $\rho_0 = \rho_+$ and the black string solution \mmm\ reduces to
\eqn\bg{\eqalign{
ds^2 & = \frac{l^2\rho^2 d\rho^2}{(\rho_+^2 - \rho^2)(\rho_-^2 - \rho^2)} - \frac{(\rho^2_+ - \rho^2_-)(\rho^2_- - \rho^2)dt^2}{4l^2\rho^2} + \frac{(\rho_+^2 - \rho_-^2)\rho^2}{4(\rho^2_+ - \rho^2)}\left(dx - \frac{\rho_-\rho_+}{l\rho^2}dt\right)^2,\cr
B & = B_{tx}dt\wedge dx, \quad B_{tx} = \frac{\rho_+^2 - \rho_-^2}{4l} = {\rm const.},\quad Q = 0,\cr
e^{2\Phi} & = g_s^2|e^{2\phi}|, \quad e^{-2\phi} = 4\left(\frac{\rho_+^2 - \rho^2}{\rho_+^2 - \rho_-^2}\right).
}
}
The scalar curvature is
\eqn\bj{R = \frac{4}{l^2}\left(\frac{\rho_+^2 - \rho_-^2}{\rho^2 - \rho_+^2}\right).
}
Thus, for an observer residing in the spacetime region $\rho < \rho_-$, \ie, the exterior region, the solution \bg\ has a singularity at $\rho = \rho_0 = \rho_+$ beyond the horizon, \ie, $\rho_0 > \rho_-$. Thus, in view of this observer, \bg\ describes a black hole. Since $x$ is compact, the solution contains closed time-likes curve in the region $\rho > \rho_0 =  \rho_+$. Thus, they are located in the interior region. We note that the $B$ field, \ie\ $B_{tx}$, is a non-zero constant. We can fix the $B$ field to zero using the gauge freedom. The Penrose diagram is given in Fig. 5.
\ifig\loc{The plot depicts the global structure of the black string solution with $\gamma = 1$.  $\rho_-$ represents the horizon. $\rho = 0$ is a causal singularity. The solution has a space-like singularity at $\rho_0 = \rho_+$. $\rho_0$ is represented by zigzag lines.}
{\epsfxsize1.8in\epsfbox{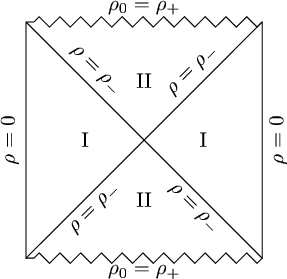}}

In this paper, we will mainly focus on the black string solutions with negative $\gamma$. In what follows, we apply duality transformation on the black string solutions described by \mmm\ to obtain other black hole solutions. See Appendix B for a review of the duality transformation.

\subsec{New black hole solutions}

We first boost the black string solutions \mmm\ along the $x$ direction. To this end, we unwrap the coordinate $x$ and write
\eqn\br{\eqalign{
t & = \hat{t}\cosh(\alpha) + l\hat{x}\sinh(\alpha),\cr
lx & = l\hat{x}\cosh(\alpha) + \hat{t}\sinh(\alpha),
}
}
where $\alpha$ is given by
\eqn\bs{\tanh(2\alpha) = \frac{2\rho_+\rho_-}{\rho_+^2 + \rho_-^2}.
}
The black string solutions now take the form
\eqn\bt{\eqalign{
d{\hat s}^2 & = \frac{l^2\rho^2d\rho^2}{(\rho^2 - \rho_+^2)(\rho^2 - \rho_-^2)} - l^{-2}e^{2\phi}(\rho^2 - \rho_+^2)d{\hat t}^2 + e^{2\phi}(\rho^2 - \rho_-^2)d{\hat x}^2,\cr
B & = B_{{\hat t}{\hat x}}d{\hat t}\wedge d{\hat x}, \cr
 B_{{\hat t}{\hat x}} & = \frac{1}{2}e^{2\phi}\left(\frac{\rho_+^2 - \rho_-^2}{l}\right)\left[\gamma - 1 - 2\left(\frac{\rho^2 - \rho_+^2}{\rho_+^2 - \rho_-^2}\right)\right],\cr
 & = \frac{1}{2l}e^{2\phi}\left[\rho_+^2(1 + \gamma) + \rho_-^2(1 - \gamma) - 2\rho^2\right],\cr
 e^{2\Phi} & = g_s^2|e^{2\phi}|,
}
}
where
\eqn\btzz{
e^{-2\phi} = \frac{\rho_+^2(1 + \gamma)^2 - \rho_-^2(1 - \gamma)^2 - 4\gamma\rho^2}{\rho_+^2 - \rho_-^2}.
}
We in particular note that there is no cross term in the metric. Therefore, in the case $\hat x$ is periodic they describe non-rotating black strings.\foot{At $\gamma = 0$, the dilaton is constant and the metric is (equivalent to) the non-rotating BTZ black hole. $\rho = \rho_-$ is a causal singularity. That is, in the case ${\hat x}$ is periodic, there are closed time-like curves in the region $0 \leq \rho < \rho_-$. At $\gamma = -1$, the coefficient of $d{\hat x}^2$ equals $\left(\rho^2_+ - \rho^2_-\right)/4$, \ie, it is a constant. One may as well prefer to refer to the solutions \bt\ as deformed black holes.}

Before we proceed further, we consider the special case in which 
\eqn\horwpartca{\gamma = {\rho_- - \rho_+ \over \rho_- + \rho_+}, \quad \rho_+ > \rho_-.
}
The black string solutions \bt\ reduce to 
\eqn\btspecial{\eqalign{
d{\hat s}^2 & = \frac{l^2\rho^2d\rho^2}{(\rho^2 - \rho_+^2)(\rho^2 - \rho_-^2)} - {\rho_+^2\over l^2(1 - \gamma)^2}\left(1 - {\rho_+^2\over \rho^2}\right)d{\hat t}^2 + {\rho_+^2\over (1 - \gamma)^2}\left(1 - {\rho_-^2\over \rho^2}\right)d{\hat x}^2,\cr
B & = B_{{\hat t}{\hat x}}d{\hat t}\wedge d{\hat x}, \quad B_{{\hat t}{\hat x}}  = {\rho^2_+\over l(1 - \gamma)^2}\left({\rho_-\rho_+\over \rho^2} - 1\right),\cr
 e^{2\Phi} & = g_s^2|e^{2\phi}| = {g_s^2\over (1 - \gamma)^2}{\rho_+^2\over \rho^2}.
}
}
These particular non-rotating solutions (with independent $\rho_\pm$ parameters) are already known in the literature and they are first obtained in \Horne. They are exact solutions of classical string theory. This family of solutions is related to \spMwww, thus also to \nnewFFF, by the Lorentz boost \br. Note $\gamma$ takes the special value \horwpartca, see also \spechorhowr\ and \ewDEFFCON. To put them into the form given in \Horne\ we introduce the new coordinates
\eqn\rednextendd{{\rho_+\over l(1 - \gamma)} {\hat t} = {l\over2} \tau, \quad {\rho_+\over (1 - \gamma)}{\hat x} = {l\over 2}\chi, \quad \rho^2 = r.
}
This gives \Horne
\eqn\btspecialNV{\eqalign{
d{\bar s}^2 & = {l^2\over 4}\left[\frac{dr^2}{r^2\left(1 - {r_+\over r}\right)\left(1 - {r_-\over r}\right)} - \left(1 - {r_+\over r}\right)d{\tau}^2 + \left(1 - {r_-\over r}\right)d{\chi}^2\right],\cr
B & = B_{{\tau}{\chi}}d{\tau}\wedge d{\chi}, \quad B_{{\tau}{\chi}}  = {l^2\over 4}\left({\sqrt{r_- r_+}\over r} - 1\right),\cr
-2\Phi & = \ln \left({|r|\over r_+}\right) + {\rm const.}, \quad {r_-\over r_+} = \left({1 + \gamma \over 1 - \gamma}\right)^2,
}
}
where $r_\pm = \rho_\pm^2$. We could more generally take $r_-$ and $r_+$ as independent parameters. That is, we could simply disregard the constraint on the ratio $r_-/r_+$, see also \horwpartca. However, this would correspond to a (fundamentally) different case. Note $\Phi_{\rm there} = -2\Phi_{\rm here}$.\foot{The constant is ${\rm const.} = \ln\left[(1 - \gamma)^2/g_s^2\right]$. It is invariant under $\gamma \to 1/\gamma$.} Note also in general $\chi$ need not be periodic (with a definite period) and $r$ can take negative values. However here we will assume, for simplicity, $\chi$ is compact and $r$ is positive. The Ricci scalar $R$ is singular at $r = 0$ and it is finite at $r = r_\pm$. It is given by $R = 2(2r(r_- + r_+) - 7r_-r_+)/(lr)^2$. We note that inside the event horizon, \ie, in the region $0 < r < r_+$, the coordinate $\tau$ is everywhere space-like. We also note that $\chi$ becomes time-like in the region $0 < r < r_-$. We will not discuss \btspecialNV\ further here. We refer to \Horne\ for more details and a comprehensive discussion.

We first note that for $\gamma \neq 0$ the $B$ field, \ie\ its component $B_{{\hat t}{\hat x}}$ \bt, is a constant at $\rho = \infty$. In general, we can use the gauge ambiguity in defining $B$ to set the constant at infinity to any arbitrary value. An equivalent Kalb-Ramond field, \ie, that gives the same $H = dB$, is
\eqn\buxzv{\eqalign{
{\tilde B}_{{\hat t}{\hat x}} & = B_{{\hat t}{\hat x}} + {\lambda(\rho_+^2 - \rho_-^2)\over 2l}, \cr
& =  \frac{1}{2l}e^{2\phi}\left[\rho_+^2(1 + \gamma) + \rho_-^2(1 - \gamma) - 2\rho^2\right] + {\lambda(\rho_+^2 - \rho_-^2)\over 2l},\cr
& = \frac{1}{2l}e^{2\phi}\left[\rho_+^2(1 + \gamma)(1 + \lambda (1 + \gamma)) + \rho_-^2(1 - \gamma)(1 - \lambda(1  - \gamma)) - 2\rho^2(1 + 2\lambda \gamma)\right],\cr
& = \frac{1}{2l}e^{2\phi}\left[\rho_+^2(1 + \lambda + \gamma(1 + 2\lambda) + \lambda\gamma^2) + \rho_-^2(1 - \lambda - \gamma(1 - 2\lambda) - \lambda\gamma^2) - 2\rho^2(1 + 2\lambda \gamma)\right].
}
}
Under $\gamma \to 1/\gamma$, $\lambda$ transforms as $\lambda \to -\gamma(1 + \gamma \lambda)$ so that not only $H$ but ${\tilde B}$ is also invariant.
 
We next periodically identify ${\hat x}$ and perform $T$-duality. This gives (in units where $\alpha' = 1$) the dilaton black hole solution described by the metric and dilaton,
\eqn\buxzv{\eqalign{
d{\tilde s}^2 & = g_{\rho\rho}d\rho^2 + {1\over g_{{\hat x}{\hat x}}}(d{\tilde x} - {\tilde B}_{{\hat t}{\hat x}}d{\hat t})^2 + g_{{\hat t}{\hat t}}d{\hat t}^2,\cr
& = \frac{l^2\rho^2d\rho^2}{(\rho^2 - \rho_+^2)(\rho^2 - \rho_-^2)} - \frac{(\rho_+^2 - \rho_-^2)(\rho^2 - \rho_+^2)}{l^2\left[\rho_+^2(1 + \gamma)^2 - \rho_-^2(1 - \gamma)^2 - 4\gamma\rho^2\right]}d{\hat t}^2 \cr
& + \left[\frac{\rho_+^2(1 + \gamma)^2 - \rho_-^2(1 - \gamma)^2 - 4\gamma\rho^2}{(\rho_+^2 - \rho_-^2)(\rho^2 - \rho_-^2)}\right]\cr
& \cdot \left\{d{\tilde x} - \frac{(\rho_+^2 - \rho_-^2)\left[\rho_+^2(1 + \gamma)(1 + \lambda (1 + \gamma)) + \rho_-^2(1 - \gamma)(1 - \lambda(1  - \gamma)) - 2\rho^2(1 + 2\lambda \gamma)\right]}{2l\left[\rho_+^2(1 + \gamma)^2 - \rho_-^2(1 - \gamma)^2 - 4\gamma\rho^2\right]}d{\hat t}\right\}^2,\cr
& = \frac{l^2\rho^2d\rho^2}{(\rho^2 - \rho_+^2)(\rho^2 - \rho_-^2)} - \frac{(\rho_+^2 - \rho_-^2)(4\lambda(1 + \gamma \lambda)\rho^2 + \rho_-^2(1 - (1 - \gamma)\lambda)^2 - \rho_+^2(1 + (1 + \gamma)\lambda)^2)}{4l^2(\rho^2 - \rho_-^2)} d{\hat t}^2\cr
& - 2\frac{\rho_-^2(1 - \gamma)(1 - (1 - \gamma)\lambda) + \rho_+^2(1 + \gamma)(1 + (1 + \gamma)\lambda) - 2(1 + 2\gamma \lambda)\rho^2}{2l(\rho^2 - \rho_-^2)}d{\tilde x}d{\hat t}\cr
& + \frac{\rho_+^2(1 + \gamma)^2 - \rho_-^2(1 - \gamma)^2 - 4\gamma \rho^2}{(\rho_+^2 - \rho_-^2)(\rho^2 - \rho_-^2)}d{\tilde x}^2,\cr
e^{-2\tilde\Phi} & = |g_s^{-2}(\rho^2 - \rho_-^2)|,
}
}
where ${\tilde x}$ is the dual coordinate. Under $\gamma \to 1/\gamma$, ${\tilde x} \to -{\tilde x}/\gamma$ and ${\hat t} \to {\hat t}/\gamma$. We note that the $\tilde B$ field, and thus in turn the gauge parameter $\lambda$, enters into the coefficient of the cross term $d{\hat t} d{\tilde x}$. The solution is invariant under $\gamma \to 1/\gamma$ because $\lambda$ transforms as $\lambda \to -\gamma(1 + \gamma \lambda)$. We will assume $0 < \lambda < -1/\gamma$ so that the solution describes a black hole. That is, we require the sign of the time component of the metric, \ie\ the coefficient of $d{\hat t}^2$, to be negative also on the section ${\tilde x} = {\rm constant}$ in the region $\rho > \rho_+$. The scalar curvature is given by
\eqn\bv{R = \frac{4(\rho_+^2 - \rho_-^2)}{l^2(\rho^2 - \rho_-^2)}.
}
Thus, the black hole solutions (labelled by the quartet $(\rho_-, \rho_+; \gamma; \lambda)$) have a curvature singularity enclosed inside their event horizons. 

We specially note that $R$ does not depend on the coupling $\gamma$. In particular for $\lambda = -1/(2\gamma)$ and, $\gamma = (\rho_- - \rho_+)/(\rho_- + \rho_+)$ the solution can be put using the coordinates \rednextendd\ into the form
\eqn\sepbhbhbh{\eqalign{d{\tilde s}^2 & = {l^2\over 4} \left[{dr^2\over\left(1 - {r_+\over r}\right)\left(1 - {r_-\over r}\right)r^2}  - \left(1 - {r_+\over r}\right)d\tau^2 + {1\over \left(1 - {r_-\over r}\right) }\left(d\chi - {\sqrt{r_- r_+}\over r}d\tau\right)^2\right],\cr
-2\tilde\Phi & = \ln(|r - r_-|) + {\rm const.},
}
}
where $r_\pm = \rho_\pm^2$ and ${\rm const.} = -2\ln g_s$. In the (extremal) limit $r_+ \to r_- := r_0$ this simplifies to 
\eqn\sepbhbhbh{\eqalign{d{\tilde s}^2 & = {l^2\over 4} \left[{dr^2\over\left(1 - {r_0\over r}\right)^2r^2}  - \left(1 - {r_0\over r}\right)d\tau^2 + {1\over \left(1 - {r_0\over r}\right) }\left(d\chi - {r_0\over r}d\tau\right)^2\right],\cr
-2\tilde\Phi & = \ln(|r - r_0|) + {\rm const.}.
}
}
We note from \bv\ that in this limit the string frame Ricci scalar $R = 0$. The Kretschmann scalar can also be shown to be zero. However, we also note that the dilaton diverges at $r = r_0$. Therefore, near $r = r_0$ the string coupling is large. In fact, in the Einstein frame, the scalar curvature depends on $r$ and it is singular at $r = r_0$. It is given by $R = -8g_s^4/(l^2(r - r_0)^2)$. Thus, in the Einstein frame, the solution \sepbhbhbh\ describes a naked singularity.\foot{The solution in the Einstein frame is \eqn\sepbhbhbhEINS{\eqalign{d{\tilde s}^2 & = {1\over 4} \cdot {k\over g_s^4} \cdot  \left[dr^2  - r^2 \left(1 - {r_0\over r}\right)^3d\tau^2 + r^2 \left(1 - {r_0\over r}\right)\left(d\chi - {r_0\over r}d\tau\right)^2\right],\cr
-2\tilde\Phi & = \ln(|r - r_0|) - 2\ln g_s, \quad g_s^2 = {8\pi G_{N}\over \kappa_0^2}, \quad k = {l^2\over l_s^2}, \quad l_s \propto l_p.
}
}
$l_p$ is Planck length.} The geometry that the string metric \sepbhbhbh\ describes, for $r > r_0$ and constant $\tau$, looks like a funnel. Thus, the strings get stretched more and more as $r$ gets closer and closer to $r_0$. Note in general $\chi$ need not be periodic with a definite period. We leave a further study of the (intriguing) solution \sepbhbhbh\ to a future work. In this paper we in general assume $\rho_+ > \rho_-$.

We note that in the case $\gamma = -1$ the black hole solution \buxzv\ reduces to
\eqn\bux{\eqalign{
d{\tilde s}^2 & = \frac{l^2\rho^2d\rho^2}{(\rho^2 - \rho_+^2)(\rho^2 - \rho_-^2)} - \frac{(\rho_+^2 - \rho_-^2)(\rho^2 - \rho_+^2)d{\hat t}^2}{4 l^2 (\rho^2 - \rho_-^2)} +\frac{4}{(\rho_+^2 - \rho_-^2)} \left[d{\tilde x} + \frac{(\rho_+^2 - \rho_-^2)(1 - 2\lambda)}{4l}d{\hat t}\right]^2,\cr
e^{-2\tilde\Phi} & = |g_s^{-2}(\rho^2 - \rho_-^2)|.
}
}
We note that for $\rho_- = 0$ it is equivalent to the solution \www\ in the sense both have a horizon at $\rho_+$ and a singularity at $\rho = \rho_- = 0$. However, here the black hole is in a spacetime with non-zero angular velocity at radial infinity and no $B$ field.

The black hole solutions given in \buxzv\ are described by the four parameters $\rho_-$, $\rho_+$, $\gamma$ and $\lambda$. Thus, in general they are not identical. To the best of my knowledge the black hole solutions \buxzv\ have not been reported elsewhere in the literature. We next find the (analogous) ADM masses and angular momenta of the solutions \ppp\ and \buxzv. The ADM masses are given in a similar way as in the BTZ black hole by the asymptotic values of the Brown-York quasi-local masses at radial infinity. The angular momenta are given by the Brown-York quasi-local angular momenta which (as we momentarily show) are independent of $\rho$ and thus are constant.

\newsec{Dilatonic rotating black hole solutions}

In this section we consider the black hole solutions \ppp\ and \buxzv. They are exact solutions of classical string theory, \ie, they have exact world-sheet CFT descriptions. The reason is essentially because we found them by applying T-dualities and coordinate transformations on $AdS_3$. More on the deformed CFTs in a future work. The solutions are again given below for convenience. The rotating black hole solution \ppp\ is given by
\eqn\cccc{\eqalign{
ds^2 & =\frac{l^2\rho^2d\rho^2}{(\rho^2 - \rho_+^2)(\rho^2 - \rho_-^2)} - \frac{1}{4}\frac{(\rho^2 - \rho_+^2)(\rho_+^2 - \rho_-^2)}{l^2\rho^2}dt^2 + \frac{1}{4}\frac{\rho^2(\rho_+^2 - \rho_-^2)}{\rho^2 - \rho_-^2}\left(dx - \frac{\rho_+\rho_-}{l\rho^2}dt\right)^2, \cr
 & = \frac{l^2\rho^2d\rho^2}{(\rho^2 - \rho_+^2)(\rho^2 - \rho_-^2)} - \frac{1}{4l^2}\frac{(\rho_+^2 - \rho_-^2)(\rho^2 - \rho_+^2 - \rho_-^2)}{\rho^2 - \rho_-^2}dt^2 - \frac{1}{2l}\frac{\rho_+\rho_-(\rho_+^2 - \rho_-^2)}{\rho^2 - \rho_-^2}dx dt\cr
 & + \frac{1}{4}\frac{\rho^2(\rho_+^2 - \rho_-^2)}{\rho^2 - \rho_-^2}dx^2, \cr
e^{-2\Phi} & = g_s^{-2}\left|e^{-2\phi}\right|, \quad e^{-2\phi} = 4\left(\frac{\rho^2 - \rho_-^2}{\rho_+^2 - \rho_-^2}\right).
}
} 
The rotating black hole solution \buxzv\ is given by
\eqn\cccz{\eqalign{
d{\tilde s}^2 & = \frac{l^2\rho^2d\rho^2}{(\rho^2 - \rho_+^2)(\rho^2 - \rho_-^2)} - \frac{(\rho_+^2 - \rho_-^2)(\rho^2 - \rho_+^2)}{l^2\left[\rho_+^2(1 + \gamma)^2 - \rho_-^2(1 - \gamma)^2 - 4\gamma\rho^2\right]}d{\hat t}^2 \cr
& + \left[\frac{\rho_+^2(1 + \gamma)^2 - \rho_-^2(1 - \gamma)^2 - 4\gamma\rho^2}{(\rho_+^2 - \rho_-^2)(\rho^2 - \rho_-^2)}\right]\cr
& \cdot \left\{d{\tilde x} - \frac{(\rho_+^2 - \rho_-^2)\left[\rho_+^2(1 + \gamma)(1 + \lambda (1 + \gamma)) + \rho_-^2(1 - \gamma)(1 - \lambda(1  - \gamma)) - 2\rho^2(1 + 2\lambda \gamma)\right]}{2l\left[\rho_+^2(1 + \gamma)^2 - \rho_-^2(1 - \gamma)^2 - 4\gamma\rho^2\right]}d{\hat t}\right\}^2,\cr
e^{-2\tilde\Phi} & = g_s^{-2}\left|e^{-2\phi}\right|, \quad e^{-2\phi} = (\rho^2 - \rho_-^2).
}
}
The parameter $\lambda$ is constrained to lie within $0 < \lambda < -{1/ \gamma}$ so the solution describes a black hole.

The scalar curvature in both cases is given by
\eqn\dddd{
R = \frac{4}{l^2}\left(\frac{\rho_+^2 - \rho_-^2}{\rho^2 - \rho_-^2}\right).
}
Thus, the inner horizon disappears and becomes singular. This is related to the inner horizon in BTZ being unstable \Chandrasekhar. However, it is important to note, as we saw earlier, that the outer horizon can also instead be singular and thus unstable upon coupling the theory to matter. Note also, however, the singularity now is in the interior region, see the case $\gamma = 1$ \bg. We hope to study this phenomenon or scenario further in the future. 

The solution \cccz\ has similar Penrose diagram to that of \cccc. It is depicted in Fig. 3.

The solutions solve the (low energy) (super)gravity action \Callan. In string frame it is simply given by setting $H = 0$ in \sgracTT,
\eqn\mmmm{
    S = \frac{1}{2\kappa^2}\int d^3 x\sqrt{-g}e^{-2\phi}\left[R + 4(\nabla\phi)^2 -2\Lambda\right].
}
The equations of motion are
\eqn\pppp{\eqalign{
    R_{\mu\nu} + 2\nabla_{\mu}\nabla_\nu \phi & = 0,\cr
    R + 4\nabla^2\phi - 4(\nabla\phi)^2 - 2\Lambda & = 0.
}
}

In the Einstein frame the low energy action is
\eqn\ssss{
S = \frac{1}{2\kappa^2}\int d^3x\sqrt{-\tilde{g}}\left[\tilde{R} - 4(\tilde{\nabla}\phi)^2 - V(\phi)\right], \quad V(\phi) = 2\Lambda e^{4\phi}, 
}
where $\tilde{g}_{ab}$ is the metric in the Einstein frame and it is given by
\eqn\tttt{
    \tilde{g}_{ab} = e^{-4\phi}g_{ab}.
}
Note under the conformal rescaling $\tilde{g}_{ab} = e^{2\alpha}g_{ab}$ the scalar curvature is not invariant. It transforms to
\eqn\yyyy{
    \tilde{R} = e^{-2\alpha}\left[R - 2(d - 1)\nabla^2\alpha - (d - 2)(d - 1)(\nabla\alpha)^2\right].
}
Thus, for $\alpha = -2\phi$ and $d = 3$ we find
\eqn\uuuu{\eqalign{
    \tilde{R} & =  e^{4\phi}\left[R - 4\nabla^2\alpha - 2(\nabla\alpha)^2\right],\cr
   & = -\frac{4g_s^4}{l^2}\frac{\left[\rho_+^2 - \rho_-^2 + 2(\rho^2 - \rho_-^2)\right]}{(\rho^2 - \rho_-^2)^3}.
}
}
The equations of motion that we get from varying the action \ssss\ are
\eqn\wwww{\eqalign{
    \tilde{R}_{ab} & = 4\tilde{\nabla}_a\phi\tilde{\nabla}_b\phi +\tilde{g}_{ab}V(\phi),\cr
    \tilde{\nabla}^2\phi & = \frac{1}{8}\frac{\partial V(\phi)}{\partial\phi}.
}
}
The first equation is the Einstein equations for the metric
\eqn\xxxx{
    \tilde{G}_{ab} := \tilde{R}_{ab} - \frac{1}{2} \tilde{g}_{ab} \tilde{R} = \kappa^2 \tilde{T}_{ab},
}
where the energy-momentum tensor is
\eqn\zzzz{
    \kappa^2 \tilde{T}_{ab} = 4\tilde{\nabla}_a \phi \tilde{\nabla}_b \phi - 2\tilde{g}_{ab} (\tilde{\nabla} \phi)^2 - \frac{1}{2}\tilde{g}_{ab}V(\phi). 
}

To compute the (analogous) ADM masses and (Komar) angular momenta we need the metrics in the Einstein frame.

In the Einstein frame, the string (frame) solution \cccc, is given by
\eqn\vvvv{\eqalign{
    ds^2 & = \frac{16 l^2}{(\rho_+^2 - \rho_-^2)^2}\frac{\rho^2(\rho^2 - \rho_-^2)}{(\rho^2 - \rho_+^2)}d\rho^2 - \frac{4}{l^2(\rho_+^2 - \rho_-^2)}(\rho^2 - \rho_-^2)(\rho^2 - \rho_+^2 - \rho_-^2)dt^2 - \frac{8\rho_+\rho_-(\rho^2 - \rho_-^2)}{l(\rho_+^2 - \rho_-^2)}dtdx \cr
   & + \frac{4\rho^2(\rho^2 - \rho_-^2)}{(\rho_+^2 - \rho_-^2)}dx^2,\cr
   & = \frac{16 l^2}{(\rho_+^2 - \rho_-^2)^2}\frac{\rho^2(\rho^2 - \rho_-^2)}{(\rho^2 - \rho_+^2)}d\rho^2 - \frac{4}{l^2(\rho_+^2 - \rho_-^2)}\frac{(\rho^2 - \rho_+^2)(\rho^2 - \rho_-^2)^2}{\rho^2}dt^2 \cr
   &+ \frac{4\rho^2(\rho^2 - \rho_-^2)}{(\rho_+^2 - \rho_-^2)}\left(dx - \frac{\rho_+\rho_-}{l\rho^2}dt\right)^2,\cr
   e^{-2\Phi} & = \left|4g_s^{-2}\left(\frac{\rho^2 - \rho_-^2}{\rho_+^2 - \rho_-^2}\right)\right|.
}
}
We can write the black hole solution \vvvv\ into the form
\eqn\vvvz{\eqalign{
    d\tilde{s}^2 & = L^{-2}(r)dr^2 - N^2(r)dt^2 + r^2\left(d\theta + Z dt\right)^2,\cr
    e^{-2\tilde\Phi} & = |g_s^{-2}f(r)|,
}
}
where $r$ is given by the equation
\eqn\rcasea{
\rho^2 = \frac{1}{2}\left(\rho_-^2 + \sqrt{\rho_-^4 + r^2(\rho_+^2 - \rho_-^2)}\right),
}
and $\theta = x$. The singularity is at $r = 0$ and the horizon is at $r = 2\rho_+$. The metric components and dilaton, \ie, $N, L, Z$ and $f$, are some known functions of $r$.

Similarly, the string frame solution \cccz, is given in the Einstein frame by
\eqn\bx{\eqalign{
d{\tilde s}^2 & = -\frac{(\rho^2 - \rho_-^2)^2(\rho_+^2 - \rho_-^2)(\rho^2 - \rho_+^2)}{l^2\left[\rho_+^2(1 + \gamma)^2 - \rho_-^2(1 - \gamma)^2 - 4\gamma \rho^2\right]}dt^2 + \frac{l^2(\rho^2 - \rho_-^2)\rho^2}{(\rho^2 - \rho_+^2)}d\rho^2 \cr
& + \left(\frac{\rho^2 - \rho_-^2}{\rho_+^2 - \rho_-^2}\right) \left[\rho_+^2(1 + \gamma)^2 - \rho_-^2(1 - \gamma)^2 - 4\gamma \rho^2\right] \cr
& \cdot \left\{dx - \frac{(\rho_+^2 - \rho_-^2)\left[\rho_+^2(1 + \gamma)(1 + \lambda (1 + \gamma)) + \rho_-^2(1 - \gamma)(1 - \lambda(1  - \gamma)) - 2\rho^2(1 + 2\lambda \gamma)\right]}{2l\left[\rho_+^2(1 + \gamma)^2 - \rho_-^2(1 - \gamma)^2 - 4\gamma\rho^2\right]}dt\right\}^2,\cr
e^{-2\tilde\Phi} & = |g_s^{-2}(\rho^2 - \rho_-^2)|.
}
}
For later convenience, we set here ${\tilde x} = x$ and ${\hat t} = t$. We can also write the black hole solution \bx\ into the form \vvvz. Here $r$ is given by the equation
\eqn\bxx{\rho^2 = \rho_-^2 - \frac{1}{8\gamma}\left[\sqrt{(\rho_+^2 - \rho_-^2)^2(1 + \gamma)^4 - 16 \gamma (\rho_+^2 - \rho_-^2) r^2} - (\rho_+^2 - \rho_-^2)(1 + \gamma)^2\right], 
}
and $\theta = x$. The singularity is at $r = 0$ and the horizon is at $r = (1 - \gamma)\sqrt{\rho_+^2 - \rho_-^2}$. The metric components and dilaton, \ie\ $N, L, Z$ and $f$, are some other known functions of $r$.

The black hole solution \vvvv, as we mentioned in the previous section, is already known in the literature and was first obtained earlier in \Chan. The authors \Chan\ studied the minimally coupled Einstein-scalar gravity in three spacetime dimensions with the scalar in an exponential potential. The solution \vvvv\ belongs to the family of black hole solutions obtained in \Chan. In \Chan\ it corresponds (using their notation) to the case $N = 1$. To write \vvvv\ in the form given in \Chan\ we need to make the change of coordinate
\eqn\ab{
    r^{2\nu} = 4\left(\frac{\rho^2 - \rho_-^2}{\rho_+^2 - \rho_-^2}\right), \quad  \nu = \frac{1}{4}.
}

We now use the Einstein frame metrics to obtain the ADM masses and angular momenta of the rotating black hole solutions \vvvv\ and \bx. We follow the Brown-York quasi-local approach \BrownY. We have reviewed the approach and summarized relevant formulae in Appendix C. Note that the dilaton is minimally coupled to gravity, see the action \ssss. It also only depends on $\rho$. Thus, it does not contribute to the ADM mass and angular momentum. See eq. (C.17) in the Appendix and the discussion in the paragraph following it.

For a $(2 + 1)$ stationary and axisymmetric spacetime of the form \Kamata
\eqn\ah{
    ds^2 = -N^2 dt^2 + \frac{1}{L^2}dr^2 + K^2(Z dt + d\phi)^2,
}
the quasi-local energy $E$, angular momentum $j$ and mass $m$ are given by (with $\kappa^2 = \pi$ in some length unit)
\eqn\aj{\eqalign{
    j & = \frac{L Z' K^3}{N},\cr
    E & = 2(\varepsilon - LK'),\cr
    m & = NE - jZ,
}
}
where $\varepsilon$ determines the zero of the energy. It is given by $\varepsilon = LK'$ for a particular solution. Primes denote derivatives with respect to $r$. $j$ agrees with the usual Komar angular momentum formula (adapted for $(2 + 1)$ dimensions). For the derivation of \aj\ and review see the Appendix. See also, {\it e.g.}, \refs{\Chan, \ \Kamata,\ \ChanK,\ \Brown}.

The angular momentum $J$ and (analogous) ADM mass $M$ are then given by 
\eqn\ajz{\eqalign{
    J & = \lim_{r \to \infty} j(r),\cr
    M & = \lim_{r\to \infty} m(r).
}
}
It is important to keep in mind that the mass $m$ is measured by the static observer whose four-velocity is $u^\mu _{(1)} = \delta^\mu_t = (1, 0, 0)$, \ie, tangent to the static Killing field $\xi^\mu = (\partial_t)^\mu$. The energy $E$ is measured by the stationary observer whose four-velocity is $u^\mu_{(2)} = (1/N)\delta^\mu_t - (Z/N)\delta^\mu_\phi$.\foot{
Note in general $u^\mu_{(2)}$ gives, in the large $r$ limit, the Kodama field \refs{\Kodama, \Kinoshita}. The Kodama vector field is given by 
\eqn\kodama{\eqalign{
k & = k^t \partial_t + k^\phi\partial_\phi,\cr
k^t & = {1\over N}LK',\cr
k^\phi & = -{Z\over N}L K' + {1\over 2K^2}\cdot {LK^3 Z'\over N}, \cr
& = -{Z\over N}L K' + {j\over 2K^2}.
}
}
In general, the Kodama field is not a Killing field. At large $r$ it is proportional to $N u^\mu_{(2)} = (\partial_t)^\mu - Z (\partial_\phi)^\mu$.} Its world-line is perpendicular to the $t = {\rm constant}$ space-like hyper-surface. Thus, the energy and mass are equal only when the shift $Z = 0$ and the lapse $N = 1$. 

In what follows we use \ajz\ to obtain the angular momentum and ADM mass of the black hole solutions. We begin with the BTZ black hole. This helps to clarify the main points and illustrate the approach.

For the rotating BTZ black hole \ze\ we have
\eqn\ak{\eqalign{
    N & = \sqrt{\frac{(\rho^2 - \rho_+^2)(\rho^2 - \rho_-^2)}{l^2\rho^2}},\cr
    L & = N, \cr
    K & = \rho, \cr
    Z & = -\frac{\rho_+\rho_-}{l\rho^2},\cr
    \Lambda & = -\frac{1}{l^2}.
}
}
We obtain the quasi-local angular momentum $j$
\eqn\al{
    j = \frac{2\rho_+\rho_-}{l}.
}
Thus, the (Komar) angular momentum $J = j$. The quasi-local energy is
\eqn\am{
    E = -2\sqrt{{\rho^2\over l^2} + \frac{J^2}{4\rho^2} -\frac{\rho^2_+ + \rho_-^2}{l^2}} + 2\varepsilon, \quad \rho \geq \rho_+.
}
A particular solution to Einstein field equations is global $AdS_3$ \zc\ or the conformally flat spacetime 
\eqn\axii{\eqalign{
    ds^2 & = {l^2d\rho^2\over \rho^2} - {\rho^2\over l^2}d\varphi^2 + \rho^2 d\psi^2, \cr
    \Lambda & = -\frac{1}{l^2}.
}
}
This solution \ie, \axii, gives $\varepsilon = \sqrt{-\Lambda \rho^2}$. The particular solution is chosen such that at constant time or $\varphi$ it asymptotically coincides or merges at large $\rho$ smoothly with the solution \ze\ \refs{\Abbott, \HawkingA}. Thus, the quasi-local mass $m$ is given by\foot{Note here the boundary $B_t$ is at large $\rho$, see Appendix C. Thus, in the expressions for $m$ and $E$, $\rho$ is assumed to be large. The same applies for the quasi-local masses and energies that appear below. Here they serve as intermediate results.}
\eqn\an{
    m = -2\left({\rho^2\over l^2} + \frac{J^2}{4\rho^2} -\frac{\rho^2_+ + \rho_-^2}{l^2}\right) + 2\sqrt{ {\rho^2\over l^2}}\sqrt{{\rho^2\over l^2} + \frac{J^2}{4\rho^2} -\frac{\rho^2_+ + \rho_-^2}{l^2}} + \frac{J^2}{2\rho^2}.
}
The (analogous) ADM mass is then 
\eqn\ao{
    M = m(\rho \to \infty) = \frac{\rho^2_+ + \rho_-^2}{l^2}.
}
Note the quasi-local mass $m$ depends on $\varepsilon$, \ie, the reference background. Different $\varepsilon$ gives different $m$. The global $AdS_3$ \zc, for instance, gives $\varepsilon = \sqrt{-\Lambda\rho^2 + 1}$. However, the ADM mass $M$ is independent of the reference solution or regularization and it is finite.

We now consider the black hole metric \vvvv. The (analogous) ADM mass and (Komar) angular momentum are obtained in \Chan, however, for convenience and completeness we obtain them here again. We have
\eqn\ap{\eqalign{
    N & = \sqrt{\frac{4}{l^2(\rho_+^2 - \rho_-^2)}\frac{(\rho^2 - \rho_+^2)(\rho^2 - \rho_-^2)^2}{\rho^2}},\cr
    L & = \sqrt{\frac{(\rho_+^2 - \rho_-^2)^2}{16 l^2}\frac{(\rho^2 - \rho_+^2)}{\rho^2(\rho^2 - \rho_-^2)}},\cr
    K & = \sqrt{\frac{4\rho^2(\rho^2 - \rho_-^2)}{(\rho_+^2 - \rho_-^2)}},\cr
    Z & = -\frac{\rho_+\rho_-}{l \rho^2}.
}
}
The quasi-local angular momentum is given by
\eqn\aq{\eqalign{
    j & = \frac{L Z' K^3}{N},\cr
    & =\frac{2\rho_+\rho_-}{l}.
}
}
Thus, the (Komar) angular momentum is 
\eqn\aqZ{
    J  =\frac{2\rho_+\rho_-}{l}.
}
The quasi-local energy is given by 
\eqn\ar{\eqalign{
    E & = 2(\varepsilon - LK'),\cr
   & = 2\varepsilon -\frac{\sqrt{\rho_+^2 - \rho_-^2}}{l}\left(\frac{2\rho^2 - \rho_-^2}{\rho^2 - \rho_-^2}\right)\sqrt{1 - \frac{\rho_+^2}{\rho^2}}.
}
}

A particular solution to the Einstein field equations is
\eqn\ax{\eqalign{
    d\tilde{s}^2 & = \frac{16 l^2}{(\rho_+^2 - \rho_-^2)^2}\rho^2 d\rho^2 - \frac{4}{l^2(\rho_+^2 - \rho_-^2)}\rho^4 dt^2 + \frac{4}{(\rho_+^2 - \rho_-^2)}\rho^4 dx^2, \cr
    e^{-2\phi} & = \frac{4}{(\rho_+^2 - \rho_-^2)}\rho^2,\cr
    \Lambda & = -\frac{2}{l^2}.
}
}
 We find
 \eqn\ay{
     \varepsilon = LK' = \frac{\sqrt{\rho_+^2 - \rho_-^2}}{l}.
 }
Thus, the quasi-local energy is
\eqn\au{
    E = \frac{\sqrt{\rho_+^2 - \rho_-^2}}{l}\left[2 -\left(\frac{2\rho^2 - \rho_-^2}{\rho^2 - \rho_-^2}\right)\sqrt{1 - \frac{\rho_+^2}{\rho^2}}\right], \quad \rho > \rho_+.
}
The quasi-local mass is given by
\eqn\av{\eqalign{
    m & = NE - jZ,\cr
   & = \frac{2\rho_-^2}{l^2} - \frac{4(\rho^2 - \rho_+^2)}{l^2} + \frac{4(\rho^2 - \rho_-^2)\sqrt{\rho^2 - \rho_+^2}}{l^2\rho^2}.
}
}
The (analogous) ADM mass is then given by
\eqn\aw{
    M = m(\rho\to \infty) = \frac{2(\rho_+^2 - \rho_-^2)}{l^2}.
}

Similarly, for the black hole solutions \bx\ we find that the angular momenta are given by  
\eqn\bxx{J = 2\left(\frac{1 - \gamma^2}{l}\right) = {Q\over g_s}, \quad -1 \leq \gamma \leq 0.
}

A particular solution to the Einstein field equations is
\eqn\axzi{\eqalign{
    d\tilde{s}^2 & = \frac{(\rho_+^2 - \rho_-^2)\rho^4}{4\gamma l^2}dt^2 -\left(\frac{4\gamma \rho^4}{\rho_+^2 - \rho_-^2}\right)\left(dx - \frac{(\rho_+^2 - \rho_-^2)(1 + 2\lambda \gamma)}{4l \gamma}dt\right)^2 + l^2 \rho^2 d\rho^2, \cr
    e^{-2\phi} & = \rho^2,\cr
    \Lambda & = -\frac{2}{l^2}.
}
}
It gives (assuming $\gamma \neq 0$)
\eqn\axxzzi{
\varepsilon = LK' = \frac{4}{l}\sqrt{\frac{-\gamma}{\rho_+^2 - \rho_-^2}}.
}
The (analogous) ADM masses are given by
\eqn\bzzx{\eqalign{
M & = 2\left(\frac{\rho^2_+ - \rho^2_-}{l^2}\right)\left[1 + \frac{(1 - \gamma^2)(1 + 2\gamma \lambda)}{4\gamma}\right], \quad -1 \leq \gamma < 0,\cr
 & = 2\left(\frac{\rho^2_+ - \rho^2_-}{l^2}\right) + \frac{2(1 - \gamma^2)}{l}{\tilde B}_{tx}(\rho = \infty), \quad -1 \leq \gamma \leq 0,
}
}
where
\eqn\batinfty{{\tilde B}_{tx}(\rho \to \infty)  = \left\{\eqalign{
 {(\rho_+^2 - \rho_-^2)(1 + 2\lambda \gamma)\over 4\gamma l} + \left({\rho_+^2 - \rho_-^2\over 4 \rho}\right)^2\left({1 - \gamma^2\over l\gamma^2}\right) + {\cal O}(1/\rho^4), \quad \gamma \neq 0\cr
 -{\rho^2\over l} + {\cal O}(\rho^0), \quad \gamma = 0.
}\right.
}
Thus, we must choose $-1/2\gamma \leq  \lambda < -1/\gamma$ for the (analogous) ADM mass to be positive. Therefore, we must either require the Kalb-Ramond field in \mmm\ to go to zero at the boundary or else we have to allow black hole solutions with negative (analogous) ADM masses. 

At $\lambda = -1/2\gamma$ we note that the Kalb-Ramond field is zero at infinity. The ADM mass and angular momentum are\foot{Here we set $\kappa_0^2 = \pi$ (in some length unit) and $g_s = 1$, see \gsgnNN.}
\eqn\specialL{M = 2\left({\rho_+^2 - \rho_-^2\over l^2}\right), \quad J = 2\left({1 - \gamma^2\over l}\right) = Q.
}
In the regime $0 < \lambda < -1/2\gamma$ the second term in the square bracket \bzzx\ is negative and thus allows having negative ADM masses. In the cases in which 
\eqn\negativM{
0 < \lambda < -{1\over 2\gamma}\left(1 + {4\gamma \over 1 - \gamma^2}\right), \quad 2 - \sqrt{5} < \gamma < 0.
}
the corresponding black hole solutions have negative masses. Note this implies for $-1 \leq \gamma \leq 2 - \sqrt{5}$ the mass is positive for any allowed value of $\lambda$.

A solution with a non-zero asymptotic Kalb-Ramond field is thus equivalent to a solution with no Kalb-Ramond field and non-zero rotation at asymptotic infinity. In particular, there is dragging, usually known as the Lense-Thirring effect, at infinity, \ie, $Z(r = \infty) \neq 0$. Thus, locally zero angular momentum observers (ZAMOs) \refs{\BardeenP, \ \Bardeen}\ located at asymptotic infinity rotate with a non-zero constant angular velocity $\Omega = Z(r = \infty)$. 

Although non-zero $\Omega$ can certainly lower (and raise) the mass, see \eg\ the general expression (C.27) of the mass in the Appendix, it does not always result in negative mass solutions. A simple example is the BTZ solution with non-zero rotation at asymptotic infinity. To allow rotation at infinity we shift $Z$ in \ap\ by a non-zero constant. We denote the constant by $\bar\lambda$ and make the change 
\eqn\examBTZ{
Z \rightarrow Z + {\bar\lambda\over l}.
}
For the solution to describe a black hole we restrict $\bar\lambda$ to the region
\eqn\examBTZZ{
 |\bar\lambda| < 1.
}
The ADM mass is given by 
\eqn\masex{
M = {1\over l^2}(\rho_+^2 + \rho_-^2 - 2\bar\lambda \rho_+\rho_-), \quad -1 < \bar\lambda < 1.
}
Thus, we note that the mass is always positive for all admissible values of $\bar\lambda$. The angular momentum does not change and it is still given by \al.

The BTZ solution with non-zero rotation $\bar\lambda$ at infinity is (or should be viewed as) the usual BTZ solution in a constant Kalb-Ramond field background.

\newsec{Discussion}

In this paper we obtained new black hole and black string solutions in the string background ${\cal A}_3$ described by the metric $g_{\mu\nu}$, Kalb-Ramond field $B_{\mu\nu}$ and dilaton $\Phi$,
\eqn\aaaXT{\eqalign{
ds^2 & = g_{\mu\nu}dx^\mu dx^\nu = l^2(d\theta^2 - e^{2\phi}\cosh^2\theta d\varphi^2 + e^{2\phi}\sinh^2\theta d\psi^2),\cr
B & = B_{01} d\varphi\wedge d\psi, \quad B_{01} = -{1\over 2}l^2e^{2\phi}(\gamma - \cosh (2\theta)),\cr
e^{2\Phi} & = g_s^2|e^{2\phi} |,
}
}
where
\eqn\bbbXT{
e^{-2\phi} = 1 + \gamma^2 - 2\gamma\cosh (2\theta).
}
The coupling $\gamma$ takes its value in the range $[-1,1]$. The spacetime ${\cal A}_3$ \aaaXT\ interpolates between $AdS_3$ and (asymptotically) flat spacetime $\IR \times S^1 \times \IR$. It is obtained by applying a sequence of $\alpha'$ exact transformations on $AdS_3$ \MAsratT. In \MAsratT\ the deformation is interpreted in general as moving holographic boundaries. The spacetime ${\cal A}_3$ is free of curvature singularity when $\gamma$ is negative. Thus, our discussion has been restricted mainly to the case $-1 \leq \gamma \leq 0$. For positive $\gamma$ the deformation generates a curvature singularity from the onset \MAsratT.

The black string solutions are described by \mmm\ which we again write below for convenience. They are obtained from ${\cal A}_3$ by applying the simple coordinate transformation \jjj\ (and making the change of variable \jjjz). They carry the axion charge per unit length $Q$ \bp. The black string solutions (labelled by the quartet $(\rho_-, \rho_+; \gamma; \lambda)$) are
\eqn\mmmXT{\eqalign{
ds^2 & = \frac{l^2\rho^2d\rho^2}{(\rho^2 - \rho_+^2)(\rho^2 - \rho_-^2)} - e^{2\phi}\frac{(\rho^2 - \rho_+^2)(\rho^2 - \rho_-^2)}{l^2\rho^2}dt^2 + e^{2\phi}\rho^2\left(dx - \frac{\rho_+\rho_-}{l\rho^2}dt\right)^2,\cr
B & = B_{tx}dt\wedge dx, \cr
B_{tx} & = \frac{1}{2l}e^{2\phi}\left[\rho_+^2(1 + \lambda + \gamma(1 + 2\lambda) + \lambda\gamma^2) + \rho_-^2(1 - \lambda - \gamma(1 - 2\lambda) - \lambda\gamma^2) - 2\rho^2(1 + 2\lambda \gamma)\right],\cr
Q & = e^{-2\Phi} {\star H} = 2\left|\frac{1 - \gamma^2}{lg_s^2}\right|,\cr
e^{2\Phi} & = g_s^2 |e^{2\phi}|,
}
}
where
\eqn\oooXT{\eqalign{
e^{-2\phi} & = \frac{\rho_+^2 (1 + \gamma)^2 - \rho_-^2(1 - \gamma)^2 - 4\gamma \rho^2}{\rho_+^2 - \rho_-^2}.
}
}
$\lambda$ is an arbitrary constant which determines the value of the $B$ field, \ie\ $B_{tx}$, at $\rho = \infty$. It does not enter in $H$. However, it has a non-trivial implication.\foot{The black string solution is invariant under the combined transformations $\rho_\pm \to \rho_\mp$, $\gamma \to -\gamma$ and $\lambda \to -\lambda$.}

The black string solution \mmmXT\ has inner and outer horizons. This is also a general feature of rotating black holes. The inner horizon is at $\rho = \rho_-$ and the outer horizon is at $\rho = \rho_+$. In this paper, we have in general assumed, unless stated otherwise, $x$ is compact and $\rho$ is positive. In the case there is no curvature singularity we identify $x$ so that there is no conical singularity. Its period does not depend on the gauge parameter $\lambda$. To the best of my knowledge the black string solutions \mmmXT\ have not been reported elsewhere in the literature.

For positive $\gamma$, \ie, $0 < \gamma \leq 1$, the solution has a ring curvature singularity at $\rho = \rho_0 \geq \rho_+$ \cccc. Therefore, the ring singularity cannot be continuously contracted or deformed to the origin without crossing the horizon(s). For $\gamma = 1$ the outer horizon becomes singular, see \bg. The singularity is behind the horizon in the interior region, \ie, $\rho_0 > \rho_-$. Note the interior region is defined in the usual way as the part of the spacetime that contains the singularity. See the Penrose diagrams in Fig. 4 and Fig. 5.

For negative $\gamma$, \ie, $-1 \leq \gamma < 0$, the solution has no singularity if the condition \bbbb,
\eqn\bbbbXT{
{\rho_-\over \rho_+} < {1 + \gamma\over 1 - \gamma},
}
is satisfied. If the above condition is not met, then the solution has a singularity at $\rho = \rho_0 \leq \rho_-$ \cccc. At $\gamma = -1$, \mmmXT\ reduces to the black hole solution \ppp
\eqn\pppXT{\eqalign{
ds^2 & =\frac{l^2\rho^2d\rho^2}{(\rho^2 - \rho_+^2)(\rho^2 - \rho_-^2)} - \frac{1}{4}\frac{(\rho^2 - \rho_+^2)(\rho_+^2 - \rho_-^2)}{l^2\rho^2}dt^2 + \frac{1}{4}\frac{\rho^2(\rho_+^2 - \rho_-^2)}{\rho^2 - \rho_-^2}\left(dx - \frac{\rho_+\rho_-}{l\rho^2}dt\right)^2,\cr
e^{2\Phi} & =  {g_s^2\over 4}\left|{\rho_+^2 - \rho_-^2 \over \rho^2 - \rho_-^2}\right|.
}
}
 It has a curvature singularity at $\rho = \rho_-$ and an event horizon at $\rho = \rho_+$. Thus, the singularity is enclosed inside the horizon. Its (analogous) ADM mass and angular momentum are given by \aqZ, \aw,
 \eqn\ADMmnja{M = {2(\rho_+^2 - \rho_-^2)\over l^2}, \quad J = {2\rho_+\rho_-\over l}.
 }
 See the Penrose diagrams in Fig. 1, Fig. 2 and Fig. 3.
  
 It follows from the cases $\gamma = \pm 1$ that both inner and outer horizons can be unstable upon coupling rotating black holes to matter. See below also for a novel case in which a ring singularity forms in between the inner and outer horizons.
 
In section three, we Lorentz boosted the black string solution \mmmXT\ along the $x$ direction with a specific rapidity \bs,
\eqn\bsXT{\tanh(2\alpha) = \frac{2\rho_+\rho_-}{\rho_+^2 + \rho_-^2}.
}
The Kalb-Ramond field is invariant under the Lorentz boost. The resulting metric however now has no cross term and thus is static, see \bt. We subsequently applied T-duality transformation along the $\hat x$ direction. This resulted in the black hole solution \buxzv,
\eqn\buxzvXT{\eqalign{
d{\tilde s}^2 & = \frac{l^2\rho^2d\rho^2}{(\rho^2 - \rho_+^2)(\rho^2 - \rho_-^2)} - \frac{(\rho_+^2 - \rho_-^2)(\rho^2 - \rho_+^2)}{l^2\left[\rho_+^2(1 + \gamma)^2 - \rho_-^2(1 - \gamma)^2 - 4\gamma\rho^2\right]}d{\hat t}^2 \cr
& + \left[\frac{\rho_+^2(1 + \gamma)^2 - \rho_-^2(1 - \gamma)^2 - 4\gamma\rho^2}{(\rho_+^2 - \rho_-^2)(\rho^2 - \rho_-^2)}\right]\cr
& \cdot \left\{d{\tilde x} - \frac{(\rho_+^2 - \rho_-^2)\left[\rho_+^2(1 + \gamma)(1 + \lambda (1 + \gamma)) + \rho_-^2(1 - \gamma)(1 - \lambda(1  - \gamma)) - 2\rho^2(1 + 2\lambda \gamma)\right]}{2l\left[\rho_+^2(1 + \gamma)^2 - \rho_-^2(1 - \gamma)^2 - 4\gamma\rho^2\right]}d{\hat t}\right\}^2,\cr
e^{-2\tilde\Phi} & = |g_s^{-2}(\rho^2 - \rho_-^2)|, \quad -1\leq \gamma \leq 0, \quad 0 < \lambda < -1/\gamma.
}
}
To the best of my knowledge this black hole solution has not been reported elsewhere in the literature. The black hole solution has a singularity at $\rho = \rho_-$ and an event horizon at $\rho = \rho_+$. See the Penrose diagram in Fig. 3. Its (analogous) ADM mass and angular momentum are given (with $g_s = 1$ and in units in which $\kappa^2 = \pi$ and $\alpha' = 1$) by \bxx, \bzzx
\eqn\bzzxXT{\eqalign{
M & = 2\left(\frac{\rho^2_+ - \rho^2_-}{l^2}\right)\left[1 + \frac{(1 - \gamma^2)(1 + 2\gamma \lambda)}{4\gamma}\right], \quad -1 \leq \gamma < 0,\cr
 & = 2\left(\frac{\rho^2_+ - \rho^2_-}{l^2}\right) + Q \cdot \Omega, \quad -1 \leq \gamma \leq 0, \cr
 J & = 2\left(\frac{1 - \gamma^2}{l}\right) = Q, \quad -1 \leq \gamma \leq 0,
}
}
where $\Omega$ is the angular velocity at infinity \batinfty
\eqn\omegainf{
\Omega = \tilde{B}_{tx}(\rho = \infty).
}

We note that the black hole depends on the gauge parameter $\lambda$. In particular we can have black holes with negative masses \negativM. The mass is positive if we choose $-1/2\gamma \leq \lambda < -1/\gamma$. Note in the case of BTZ solutions, those solutions with negative masses do not correspond to black holes, see \bl\ and \conical. We also note that the deformation slows down the rotation \MAsratT. That is, the norm of $J$ decreases as we increase the norm of $\gamma$. The axion charge (per unit length) $Q$ of the black string (with $g_s = 1$) equals the angular momentum $J$ of the black hole. This is consistent with the work \HorneS.

At $\gamma = -1$ we note that $J = 0$ and the mass $M$ is independent of the gauge parameter $\lambda$. $M$ is determined only by the radii $\rho_+$ and $\rho_-$. At $\lambda = -1/2\gamma$, the angular velocity at infinity is zero, \ie, $\Omega = 0$. In this case the ADM mass and angular momentum reduce to \specialL
\eqn\specialLL{M = 2\left({\rho_+^2 - \rho_-^2\over l^2}\right), \quad J = 2\left({1 - \gamma^2\over l}\right) = Q.
}

In this paper we also noted that a solution with a non-zero asymptotic Kalb-Ramond field is equivalent to a solution with no Kalb-Ramond field and non-zero rotation at asymptotic infinity. In general, non-zero rotation allows negative mass solutions.

In an upcoming work we study the thermodynamics and partition functions of the black hole and black string solutions. We also study the causal structure of the solutions in detail.

In this paper we mainly presented black hole and black string solutions. In a future work we hope to study in greater detail the marginal operator that generates the deformation on the world-sheet.

We also hope to study other string solutions by gluing those solutions of positive $\gamma$ with negative $\gamma$. We give below an example which is obtained by gluing \ppp\ and \bg. It is convenient to introduce first the parameters $0 < {\bar \rho}_- < {\bar \rho}_+ = \rho_- < \rho_+$. We define $\delta = (\rho_+ - {\bar \rho}_-)/\rho_-$. We assume $\rho_+ < \sqrt{2}\rho_-$. We furthermore assume ${\bar \rho}_- = \sqrt{2\rho_-^2 - \rho_+^2}$.

For $\rho \leq {\bar \rho}_+ = \rho_-$, the solution is given by
\eqn\newPES{
   \eqalign{
       ds^2 = & {l^2\rho^2d\rho^2\over ({\bar\rho}_+^2 - \rho^2)({\bar\rho}_-^2 - \rho^2)} - {({\bar\rho}_+^2 - {\bar\rho}_-^2)({\bar\rho}_-^2 - \rho^2)dt^2\over 4l^2\rho^2} + {({\bar \rho}_+^2 - {\bar\rho}_-^2)\rho^2\over 4({\bar \rho}_+^2 - \rho^2)}\left[dx - {1\over l}\left(\delta + {{\bar \rho_+}{\bar \rho_-}\over \rho^2}\right)dt\right]^2,\cr
       B_{tx} & = {{\bar \rho}_+^2 - {\bar\rho}_-^2\over 4l} = {\rm const.}, \quad Q = 0,\cr
       e^{2\Phi} & = {g_s^2\over 4}\left({\bar \rho}_+^2 - {\bar \rho}_-^2\over {\bar \rho}_+^2 - \rho^2\right).
}
}
For $\rho \geq {\bar \rho}_+ = \rho_-$, the solution is given by
\eqn\newPESPP{\eqalign{
ds^2 & =\frac{l^2\rho^2d\rho^2}{(\rho^2 - \rho_+^2)(\rho^2 - \rho_-^2)} - \frac{1}{4}\frac{(\rho^2 - \rho_+^2)(\rho_+^2 - \rho_-^2)}{l^2\rho^2}dt^2 + \frac{1}{4}\frac{\rho^2(\rho_+^2 - \rho_-^2)}{\rho^2 - \rho_-^2}\left(dx - \frac{\rho_+\rho_-}{l\rho^2}dt\right)^2,\cr
B_{tx} & = -\left(\frac{\rho_+^2 - \rho_-^2}{4l}\right) = {\rm const.}, \quad Q = 0,\cr
e^{2\Phi} & = {g_s^2\over 4}\left(\rho_+^2 - \rho_-^2\over \rho^2 - \rho_-^2\right).
}
}
The scalar curvature of the solution (which is described by \newPES\ and \newPESPP) is $R = -4\xi^2\cdot\left(\rho_+^2/l^2\right)\cdot\left(1/|\rho^2 - \rho^2_-|\right)$, where $\xi^2 = 1 - \left(\rho_-/\rho_+\right)^2$. The Kretschmann scalar is $K = R^2$.\foot{In general, the equality $K = R^2$ is true only in special cases.} Thus, the solution has the peculiar feature that it contains a ring curvature singularity at $\rho = {\bar \rho}_+ = \rho_-$ in between the inner and outer horizons. The inner horizon is at $\rho = {\bar \rho}_-$. The outer horizon is at $\rho = \rho_+$. Note they are not related by sign reflection, \ie\ ${\bar \rho}_- \neq -\rho_+$. The Kalb-Ramond field is a constant (field). It can be fixed to zero. The Penrose diagram of the solution is given in Fig. 6. It is obtained by combining Fig. 3 and Fig. 5.
\ifig\loc{The plot depicts the global structure of the string solution described by \newPES\ and \newPESPP. The solution has a singularity at $\rho = \rho_- = {\bar\rho}_+$. It is represented by the zigzag lines. $\rho = 0$ is a causal singularity. That is, the maximal analytic extension behind the surface $\rho = 0$ contains closed time-like curves.}
{\epsfxsize2.4in\epsfbox{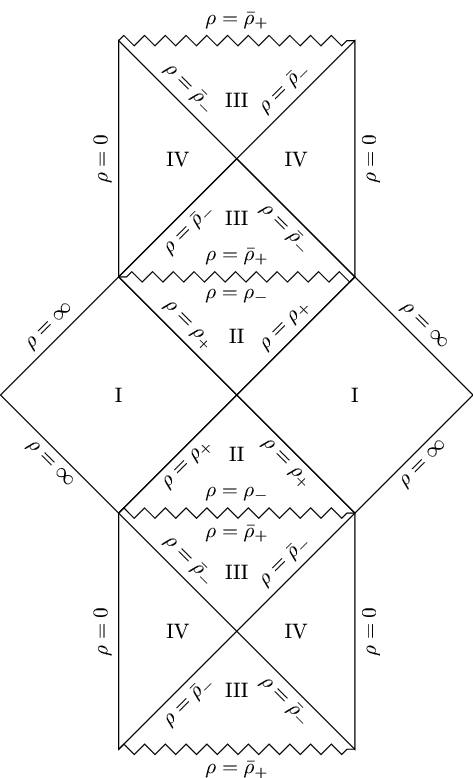}}

Note that usually the diagram ends at either side of $\rho = {\bar\rho}_+ = \rho_-$, however here, since there are no closed time-like curves, the diagram extends outward to infinity and inward to zero. Therefore, the maximal extension simply extends infinitely in both vertical directions.

We expect the black hole solution (described by \newPES\ and \newPESPP) to have a generalization to $(3 + 1)$ dimensions with similar feature. That is, with a curvature singularity in between the inner and outer horizons. We hope to study the corrections in $g_s$ and their roles in resolving the singularity in a future work. In relation to this, we also would like to better understand the solution \sepbhbhbh,
\eqn\sepbhbhbhAGAGA{\eqalign{d{\tilde s}^2 & = {l^2\over 4} \left[{dr^2\over\left(1 - {r_0\over r}\right)^2r^2}  - \left(1 - {r_0\over r}\right)d\tau^2 + {1\over \left(1 - {r_0\over r}\right) }\left(d\chi - {r_0\over r}d\tau\right)^2\right],\cr
-2\tilde\Phi & = \ln(|r - r_0|) + {\rm const.}.
}
}
Its Ricci scalar is zero.

\bigskip\bigskip

\noindent{\bf Acknowledgements:} This work is supported by the Department of Atomic Energy under project no. RTI4001.

\appendix{A}{Invariant metric on $AdS_3$}

The (invariant) metric of the Lie algebra $su(1, 1) \cong sl(2, \IR)$ is given by the Killing form $ds^2 = \tr \left(g^{-1}dg g^{-1} dg\right)$ (up to a constant factor), where $g^{-1}dg$ is the (left invariant) Maurer-Cartan one form and $g$ is $SU(1, 1) \cong SL(2, R)$ group element.

A convenient choice of basis for $su(1,1)$ is $a_1 = -{1\over 2}\sigma_1, a_2 = -{1\over 2}\sigma_2, a_3 = {i\over 2}\sigma_3$, where $\sigma_1$, $\sigma_2$ and $\sigma_3$ are the Pauli spin matrices. They are given by
\eqn\pauliiiTT{\sigma_1 = \pmatrix{ 0 & 1\cr
1 & 0}, \quad \sigma_2 = \pmatrix{ 0 & -i\cr
i & 0}, \quad \sigma_3 = \pmatrix{ 1 & 0\cr
0 & -1}.
}
We parametrize (using the Cartan decomposition) the group element $g \in SU(1,1)$ as
\eqn\slrpara{\eqalign{g(\theta, \varphi, \psi) & = e^{\left(\varphi + \psi \right)a_3}e^{2\theta a_1}e^{\left(\varphi - \psi \right)a_3},\cr
& = e^{i\left(\varphi + \psi \over 2\right)\sigma_3}e^{-\theta\sigma_1}e^{i\left(\varphi - \psi \over 2\right)\sigma_3}, \cr
& = \pmatrix{
e^{i\varphi}\cosh\theta & -e^{i\psi}\sinh\theta\cr
-e^{-i\psi}\sinh\theta & e^{-i\varphi}\cosh\theta
},
}
}
here $0 \leq \theta < \infty, 0\leq \varphi < 2\pi, 0 \leq \psi < 2\pi$. This choice gives the invariant metric 
\eqn\relgga{ds^2 = d\theta^2 - \cosh^2\theta d\varphi^2 + \sinh^2\theta d\psi^2.
}
On the universal cover $\varphi \in \IR$.

However, as we see in section two, it is the real form of the Lie algebra $su(1,1)$ (\ie, now the field is $\IR$ however the basis can be complex) or simply the lie algebra $sl(2, R)$ which is in particular relevant for our discussion in relation to the exterior region of the BTZ black hole. Therefore, we obtain below the metric using $sl(2,\IR)$. A convenient choice of basis is $b_1 = -{1\over 2}\sigma_1, b_2 = {i\over 2}\sigma_2, b_3 = -{1\over 2}\sigma_3$. The analogous parametrization is
\eqn\slrparaot{\eqalign{g(\bar\theta, \bar\varphi, \bar\psi) & = e^{\left(\bar\varphi + \bar\psi \right)b_3}e^{2\bar\theta b_1}e^{\left(\bar\varphi - \bar\psi \right)b_3},\cr
& = e^{-\left(\bar\varphi + \bar\psi \over 2\right)\sigma_3}e^{-\bar\theta\sigma_1}e^{-\left(\bar\varphi - \bar\psi \over 2\right)\sigma_3},\cr
&  = \pmatrix{
e^{-\bar\varphi}\cosh\bar\theta & -e^{-\bar\psi}\sinh\bar\theta\cr
-e^{\bar\psi}\sinh\bar\theta & e^{\bar\varphi}\cosh\bar\theta
},
}
}
here $\bar\theta \in \IR, \bar\varphi \in \IR, \bar\psi \in \IR$. In this choice the (invariant) metric is given by
\eqn\relgg{d{\bar s}^2 = d\bar\theta^2 + \cosh^2\bar\theta d\bar\varphi^2 - \sinh^2\bar\theta d\bar\psi^2.
}
The metric \relgg\ can be equivalently obtained from \relgga\ formally by the analytic continuation
\eqn\analyticggg{\varphi = i\bar\varphi, \quad \psi = i\bar\psi.
}
We simple replace $\theta$ with ${\bar \theta} \in \IR$.

\appendix{B}{Bosonic T-duality}

We briefly review T-duality in Type II bosonic string theory. See \refs{\Buscher\BuscherB-\Rocek, \ \BerkovitsB} for detailed discussions. See also Appendix A in \MAsratT.

Consider the world-sheet sigma model action
\eqn\zzza{S := \int d^2 z{\cal L}, \quad {\cal L} = \partial X^m \Sigma_{mn}(X) \bar\partial X^n, \quad \Sigma_{mn} := G_{mn} + B_{mn},
}
where $G_{mn} = G_{nm}$ are the components of the metric $G$ and $B_{mn} = -B_{mn}$ are the components of the Kalb-Ramond two-form $B$. $z = \tau + \sigma$ and ${\bar z} = \tau - \sigma$ are the world-sheet coordinates. We omitted a term linear in the dilaton field. The dilaton $\Phi$ enters at the quantum level \refs{\FradkinT, \FDavidMM}.

We assume the background fields $G_{mn}$, $B_{mn}$ and $\Phi$ are independent of $X^1$. Thus, the transformation
\eqn\zzzb{X^1 \to X^1 + c,
}
is a symmetry of the action, where $c$ is a constant. Therefore, we can couple the theory to a gauge field $A_a$ on the world-sheet. The gauged Lagrangian density is
\eqn\zzzc{{\cal L} = \partial X^m \Sigma_{mn} \bar\partial X^n + \partial X^m\Sigma_{m1} {\bar A} + A \Sigma_{1n} \bar\partial X^n + A\Sigma_{11} {\bar A} + \chi  (\partial {\bar A} - \bar\partial A), 
}
where the field $\chi$ is a Lagrange multiplier.

We assume $X^1$ is periodic and compact, \ie, at fixed world-sheet time it parametrizes a circle. Thus, the symmetry \zzzb\ is a global $U(1)$ (zero-form) symmetry. We fix the gauge by assuming $X^1 = {\rm const.}$. This gives
\eqn\zzzd{S = \int d^2z {\cal L}, \quad {\cal L} = \partial X^{\hat m} \Sigma_{{\hat m}{\hat n}} \bar\partial X^{\hat n} + \partial X^{\hat m}\Sigma_{{\hat m}1} {\bar A} + A \Sigma_{1{\hat n}} \bar\partial X^{\hat n} + A\Sigma_{11}{\bar A} + \chi  (\partial {\bar A} -  \bar \partial A),
}
 where $\hat m$ takes all values of $m$ except $m = 1$. Integrating out the field $\chi$, \ie, eliminating it using its equation of motion, gives back \zzza. In general a one form on $S^1$ is not exact. However, since $X^1$ is compact and thus it winds around in either directions integer times, we can still write the gauge potential on $S^1$ as the gradient of a compact scalar $A = \partial\chi^1, {\bar A} = {\bar \partial}\chi^1$ and make the identification $\chi^1 \sim X^1$. This will get us back to the original action.
 
 If we instead perform integration by parts, it leads to
\eqn\zzze{S' = \int d^2z {\cal L}', \quad {\cal L}' = \partial X^{\hat m} \Sigma_{{\hat m}{\hat n}} \bar\partial X^{\hat n} + A \Sigma_{11} {\bar A} - \left(\partial\chi - \partial X^{\hat m}\Sigma_{{\hat m}1} \right){\bar A} + A\left(\Sigma_{1{\hat n}} \bar\partial X^{\hat n} + \bar \partial \chi\right).
}
Integrating out the gauge fields $A$ and $\bar A$ using their equations of motion gives the T-dual Lagrangian density
\eqn\zzzf{{\cal L}' = \partial X^{\hat m} \sigma_{{\hat m}{\hat n}} \bar\partial X^{\hat n} + \partial \chi \sigma_{\chi {\hat m}} {\bar\partial  X}^{\hat m} + \partial X^{\hat m} \sigma_{{\hat m}\chi}\bar\partial\chi + \partial \chi  \sigma_{\chi\chi}  \bar \partial \chi,
}
where
\eqn\zzzg{ \sigma_{\chi\chi} = {1\over \Sigma_{11}}, \quad \sigma_{\chi\hat m} =  {\Sigma_{1\hat m}\over \Sigma_{11}}, \quad \sigma_{{\hat n}\chi} =  -{\Sigma_{{\hat n} 1}\over \Sigma_{11}}, \quad \sigma_{{\hat m}{\hat n}} = \Sigma_{{\hat m}{\hat n}} - {\Sigma_{{\hat m}1}\Sigma_{1{\hat n}}\over \Sigma_{11}}.
}
The Jacobian factor introduced in the path integral measure by the change of variables leads to a shift of the dilaton. The new dilaton $\phi$ is
\eqn\zzzh{\phi = \Phi -{1\over 2}\log \left(\Sigma_{11}\right), \quad \Sigma_{11} = G_{11}.
}
The new background fields are arranged as
\eqn\zzzi{\sigma_{{\hat m}{\hat n}} = g_{{\hat m}{\hat n}} + b_{{\hat m}{\hat n}}, \quad g_{{\hat m}{\hat n}} = g_{{\hat n}{\hat m}}, \quad b_{{\hat m}{\hat n}} = -b_{{\hat n}{\hat m}},
}
and similar expressions for $\sigma_{\chi{\hat n}},\ \sigma_{{\hat n}\chi}, \ \sigma_{\chi\chi}$. Thus, the components of the new metric $g$ and two-form $b$ field are given in terms of $G_{mn}$ and $B_{mn}$ by the relations
\eqn\zzzj{g_{\chi\chi} = {1\over G_{11}}, \quad g_{\chi{\hat m}} = {B_{1{\hat m}}\over G_{11}}, \quad g_{{\hat n}\chi} = -{B_{{\hat n} 1}\over G_{11}}, \quad g_{{\hat m}{\hat n}} = G_{{\hat m}{\hat n}} -{1\over G_{11}}\left(G_{{\hat m}1}G_{1{\hat n}} + B_{{\hat m}1}B_{1{\hat n}}   \right),
}
and
\eqn\zzzk{b_{\chi{\hat m}} = {G_{1{\hat m}}\over G_{11}}, \quad b_{{\hat n}\chi} = -{G_{{\hat n} 1}\over G_{11}}, \quad b_{{\hat m}{\hat n}} = B_{{\hat m}{\hat n}} -{1\over G_{11}}\left(G_{{\hat m}1}B_{1{\hat n}} + B_{{\hat m}1}G_{1{\hat n}}   \right).
}
The relations \zzzh, \zzzj\ and \zzzk\ are also known as the Buscher rules. From the world-sheet perspective, T-duality is an exact equivalence of CFTs.

\appendix{C}{The Brown-York quasi-local energy and mass}

In this appendix we review the Brown-York quasi-local energy and mass \BrownY. We primarily follow the work \BrownY, but see also \HawkingA\ for a related insightful discussion. We specialize the discussion for definiteness to $(2 + 1)$ dimensions. We furthermore restrict ourselves to axially symmetric stationary solutions. Therefore, we can directly use the formulae in section four of the paper.

The Brown-York definition of energy and mass is based on the Hamilton-Jacobi formulation of general relativity. It requires foliating the spacetime into a one-parameter family of space-like hyper-surfaces and thus requires defining a time direction. It also requires adding the Gibbons-Hawking-York (GHY) boundary term to the usual action for the gravitational field \refs{\YorkBC, \Gibbons}. The boundary term ensures that only the metric is fixed at the boundary. It avoids the need to fix the first derivatives of the metric. 

The general metric (relevant for our purpose) is of the form
\eqn\threddd{ds^2 = -A^2dt^2 + {1\over B^2}dr^2 + C^2(d\theta + Zdt)^2,
}
where $A, B, C$ and $Z$ are functions of only the radial coordinate $r$. The coordinate $\theta$ is an angle variable and it has $2\pi$ periodicity. The spacetime contains no inner boundaries. It can have event horizons \HawkingA.\foot{In the case the spacetime contains other null boundaries we include appropriate terms in the action functional that contribute on these boundaries. This will not concern us in this paper. We refer to \Chandrasekaran\ and references therein for such cases.} We assume the metric is minimally coupled to a scalar. Therefore, the scalar action contains no derivatives of the metric. We also assume the scalar depends only on $r$.

Let $u^\mu$ denote the future pointing time-like unit normal to the $t = {\rm constant}$ two-space denoted by $\Sigma_t$. $t$ labels the two-spaces $\Sigma_t$ and it defines the time direction or coordinate. This particular slicing is the usual ADM slicing of spacetime \Arnowitt. We denote the (intrinsic) metric on $\Sigma_t$ by $h_{ij}$.\foot{For convenience, here and below we use lower latin letters $i = t, r, \theta$ to denote the intrinsic metrics and we use greek letters $\mu = t, r, \theta$ to denote the induced metrics.} We denote the boundary of $\Sigma_t$ by $B_t$. The world-line or history of $B_t$ is denoted $^2 B$.  Let $n^\mu$ is the outward pointing space-like unit normal to the two-boundary $^2B$.  For simplicity, we assume on the time-like boundary $^2B$ that $n^\mu u_\mu|_{^2B} = 0$. That is, the surface $\Sigma_t$ meets the boundary $^2B$ orthogonally. This is not however essential or necessary for the analysis below \refs{\Hayward}. This is in part because we are ultimately interested in the case $^2B$ is at radial infinity where the spacetime \threddd\ is (conformally) flat. We denote the metric on $^2B$ by $\gamma_{ij}$. We denote the metric on $B_t$ which is the intersection of $^2B$ and $\Sigma_t$ by $\sigma_{ij}$.

The time-like and space-like normal vector fields are
\eqn\nnuu{\eqalign{u_\mu = -A\delta^t_\mu, \quad u^\mu = g^{\mu\nu}u_\nu = {1\over A}\delta^\mu_t - {Z\over A}\delta^\mu_\theta,\cr
n_\mu = {1\over B}\delta^r_\mu, \quad n^\mu = g^{\mu\nu}n_\nu = B\delta^\mu_r.
}
}

The induced metrics are
\eqn\indmeh{\eqalign{
h_{\mu\nu} & = g_{\mu\nu} + u_\mu u_\nu, \cr
& = g_{\mu\nu} + A^2\delta^t_\mu \delta^t_\nu,\cr
\gamma_{\mu\nu} & = g_{\mu\nu} - n_\mu n_\nu,\cr
& = g_{\mu\nu} - {1\over B^2}\delta^r_\mu \delta^r_\nu,\cr
\sigma_{\mu\nu} & = g_{\mu\nu} + u_\mu u_\nu - n_\mu n_\nu,\cr
& = g_{\mu\nu} + A^2\delta^t_\mu \delta^t_\nu - {1\over B^2}\delta^r_\mu \delta^r_\nu.
}
} 

The extrinsic curvature to the surface $\Sigma_t$ is
\eqn\kakakext{\eqalign{K_{\mu\nu} & = -h^\alpha_\mu\nabla_\alpha u_\nu,\cr
& = {C^2 Z (\partial_rZ) \over 2A}(\delta^t_\nu\delta^r_\mu + \delta^t_\mu\delta^r_\nu) + {1\over 2A}C^2 (\partial_r Z)(\delta^\theta_\nu\delta^r_\mu + \delta^\theta_\mu\delta^r_\nu),\cr
& = {C^2 Z (\partial_rZ) \over A}\delta^t_{(\nu}\delta^r_{\mu)} + {1\over A}C^2 (\partial_r Z)\delta^\theta_{(\nu}\delta^r_{\mu)}.
}
}

The extrinsic curvature to the boundary $^2B$ is
\eqn\skakaex{
\eqalign{\Theta_{\mu\nu} = & -\gamma^\alpha_\mu\nabla_\alpha n_\nu,\cr
= & -BC\left[C(\partial_r Z) + 2 Z(\partial_r C)\right]\delta^t_{(\mu}\delta^\theta_{\nu)} - BC(\partial_r C)\delta^\theta_\mu\delta^\theta_\nu \cr
& + B\left[A(\partial_r A) - C^2 Z(\partial_r Z) - Z^2C(\partial_r C)\right]\delta^t_\mu\delta^t_\nu.
}
}

The extrinsic curvature to the boundary $B_t$ is
\eqn\skssdkx{
\eqalign{k_{\mu\nu} & = -\sigma^\alpha_\mu\nabla_\alpha n_\nu,\cr
& = -BC(\partial_r C) \delta^\theta_\mu \delta^\theta_\nu - Z^2 BC(\partial_r C) \delta^t_\mu \delta^t_\nu - 2 ZB C(\partial_r C) \delta^\theta_{(\mu}\delta^t_{\nu)},\cr
& = \sigma^\alpha_\mu \sigma^\beta_\nu \Theta_{\alpha\beta}.
}
}
The trace $k$ is
\eqn\trkkkk{k = \sigma^{\mu\nu}k_{\mu\nu} = -BC(\partial_r C) \sigma^{\theta\theta} = -{B(\partial_r C)\over C}.
}

We denote the canonical momenta on $^2B$ conjugate to $\gamma_{ij}$ by $\pi^{ij}$. The momenta $\pi^{ij}$ contain no scalar contribution since the scalar field is coupled to gravity minimally. Thus, they are purely gravitational. The gravitational canonical momenta are given by
\eqn\momaa{\eqalign{
\pi^{ij} & = -{1\over 2\kappa^2}\sqrt{-\gamma}\left(\Theta\gamma^{ij} - \Theta^{ij}\right),
}
}
where $\Theta_{ij} = \gamma^\mu_i\gamma^\nu_j\Theta_{\mu\nu}$ and $\Theta = g^{\mu\nu}\Theta_{\mu\nu} = \gamma^{ij}\Theta_{ij}$ is the trace. In general, $\pi^{ij}$ defines a total energy-momentum tensor on the surface $^2B$. The surface energy-momentum tensor $\tau_{ij}$ is
\eqn\surftt{
\tau^{ij} := {2\over \sqrt{-\gamma}} {\delta S_{\rm on-shell}\over \delta \gamma_{ij}} = {2\over \sqrt{-\gamma}}\pi^{ij},
}
where $S_{\rm on-shell}$ is the on-shell action functional. It satisfies the momentum constraints
\eqn\rewrr{
{\cal D}_i\tau^{ij} = -T^{\mu\nu}n_{\mu}\gamma^j_\nu \equiv -T^{nj},
}
where $T^{\mu\nu}$ is the matter energy-momentum tensor and ${\cal D}_i$ is the covariant derivative compatible with the metric $\gamma_{ij}$, \ie, ${\cal D}_i\gamma^{jk} = 0$. It is defined by projecting $\nabla_\mu$ onto $^2B$, \ie, ${\cal D}_i = \gamma_i^\lambda \nabla_\lambda$. The proper surface energy density $\varepsilon$ is
\eqn\suene{\eqalign{
\varepsilon & \equiv u_i u_j \tau^{ij}  = {1\over \kappa^2} k,\cr
& =  {1\over \kappa^2}\cdot -{B(\partial_r C)\over C}.
}
}
The proper surface momentum density $j_i$ is
\eqn\momsudd{\eqalign{j_k & \equiv -\sigma_{k i}u_j\tau^{ij}   = {1\over \kappa^2} \sigma^i_k n^j K_{i j},\cr
& = {BC^2(\partial_r Z)\over 2\kappa^2 A}\left(\delta^\theta_k + Z \delta^t_k\right).
}
}

For a minimally coupled scalar $\phi$ with action \ssss\ the energy momentum tensor is given by \zzzz,
\eqn\scaltt{\kappa^2T_{\mu\nu} = 4\nabla_\mu \phi \nabla_\nu\phi - 2g_{\mu\nu}\left(\nabla \phi\right)^2 - {1\over 2}g_{\mu\nu}V(\phi).
}
Thus, the projection onto $^2B$ is
\eqn\projjjnng{\eqalign{T^{nj} & = T^{\mu\nu}n_{\mu} \gamma^j_{\nu},\cr
& = {1\over B}\left(T^{rt}\delta^j_t + T^{r\theta} \delta^j_{\theta} \right).
}
}
We have (since $\gamma^j_r = 0$)
\eqn\retttre{\eqalign{
T^{rt} & = {B^2\over N^2}\left(-T_{rt} + Z T_{r\theta}\right),\cr
T^{r\phi} & = {B^2\over N^2}\left[ZT_{rt} + \left({A^2\over C^2} - Z^2\right)T_{r\theta}\right].
}
}
In the case the scalar $\phi$ is a function of only $r$, which is the case in this paper, and the metric is of the form \threddd, we have
\eqn\condttnn{T_{rt} = 0, \quad T_{r\theta} = 0.
}
Therefore, 
\eqn\condttnnNMNM{T^{nj} = 0.
}

Let $\xi^i$ denote a Killing vector field associated with an isometry of the boundary two-metric $\gamma_{ij}$. Thus, it solves 
\eqn\kill{{\cal D}_{(i}\xi_{j)} = 0.
}
In the case $T^{ni} = 0$ (or $T^{ni}\xi_i = 0$) on the surface $^2B$, which is the case in this paper (see \condttnnNMNM), it follows from \rewrr\ that the quantity, 
\eqn\chargsim{Q_\xi(B_t) =  -\oint_{{\Sigma_{t}\ \cap \ ^2B}} d\theta\sqrt{\sigma}  \left(u_i\tau^{ij}\xi_j\right) = \oint_{{\Sigma_{t}\ \cap \ ^2B}} d\theta \sqrt{\sigma}  \left(\varepsilon u^i +j^i\right) \xi_i,
}
is conserved, \ie, $Q_\xi(B_{t_0)} - Q_\xi(B_{t_1}) = -\int_{^2B} d^2x \sqrt{-\gamma} T^{ni}\xi_i = 0$.

We now give the conserved charges in terms of the components of the metric. The energy associated with the Killing field $\xi^i = u^i =  \gamma_i^\mu u_\mu$\foot{Note ${\cal D}_{(i}\xi_{j)} = \gamma_{(i}^\lambda \nabla_\lambda \gamma_{j)}^\mu u_\mu = 0$.} is given by
\eqn\eneneby{E = -Q_{u^i} = \oint_{B_t} d\theta \sqrt{\sigma}\varepsilon = -{2\pi\over \kappa^2} B\partial_r C.
}
It is important to note that the energy equals the Hamiltonian only if the angular velocity $Z$ at radial infinity or boundary $B_t$ is zero (assuming $j_\theta \neq 0$) and the lapse $A$ is unity.\foot{The Hamiltonian upon using the equations of motion is simply a boundary term \Regge. It is given by $H = \oint_{B_t}d\theta \sqrt{\sigma}\left(A\varepsilon - Z j_\theta\right)$.} The angular momentum associated with the Killing field $\xi^i = \left(\partial_\theta\right)^i = \delta^i_\theta$ is
\eqn\jjangcn{j = Q_{(\partial_\theta)^i} = \oint_{B_t}d\theta\sqrt{\sigma}j_\theta = {\pi\over \kappa^2}{BC^3\partial_r Z\over A}.
}
The mass associated with the Killing field $\xi^i = \left(\partial_t\right)^i = \delta^i_t$ is given by
\eqn\mamamasss{m = -Q_{(\partial_t)^i} = -\oint_{B_t}d\theta\sqrt{\sigma}(\varepsilon u_t + j_t) = AE - Z j.
}
Note the mass corresponds as in asymptotically flat spacetimes to the Hamiltonian.

In gravity theories physical quantities such as energy are defined only with respect to a reference. Thus, we must specify a reference spacetime ${\cal M}_0(g_0, \phi_0; \Lambda)$, where $g_0$ denotes the metric and $\phi_0$ collectively denotes the matter fields on ${\cal M}_0$. $\Lambda$ is a possible cosmological constant.

The reference spacetime ${\cal M}_0(g_0, \phi_0; \Lambda)$ is a particular solution to Einstein equations. It is chosen such that the deviations or fluctuations
\eqn\condmecondme{\eqalign{h_{\mu\nu} := g_{\mu \nu} -( g_0)_{\mu\nu},\cr
\chi := \phi - \phi_0,
}
}
vanish smoothly at the boundary $B_t$ \Abbott. Note $B_t$ is the boundary of the $t = {\rm constant}$ spacetime slice. In the case ${\cal M}(g, \phi; \Lambda)$ is asymptotically flat spacetime, the reference spacetime ${\cal M}_0(g_0, \phi_0; \Lambda)$ is usually taken to be the Minkowski spacetime. In the paper in all the cases we study the reference metrics are conformally flat spacetimes.

We take a reference metric $g_0$ of the form
\eqn\thredref{ds^2_0 = -A_0^2dt^2 + {1\over B^2_0}dr^2 + C^2_0(d\theta + Z_0dt)^2.
}
We will assume $A_0, B_0, C_0$ and $Z_0$ are functions of only $r$. The scalar field depends only on $r$ and it is coupled to the metric minimally. At the boundary $B_t$, $h_{\mu\nu} = 0$ and $\chi = 0$ \condmecondme. That is, $A(r^\star) = A_0(r^\star), B(r^\star) = B_0(r^\star), C(r^\star) = C_0(r^\star), Z(r^\star) = Z_0(r^\star)$ and $\phi(r^\star) = \phi_0(r^\star)$, where $r^\star$ is the location of $B_t$.

The reference metrics $g_0$ that we consider in this paper have $Z_0 = {\rm constant}$. Accordingly, since $\partial_r Z_0 = 0$, the (quasi-local) angular momentum remains unchanged. It is still given by \jjangcn. The (quasi-local) energy is given by\foot{In general it is quasi-local in the sense that it is local only on the (hyper-)surface $\Sigma_t$, \ie, the $t = {\rm constant}$ spacetime slice. It follows, in $(2 + 1)$ dimensions, it is defined as a local quantity.}
\eqn\modifddeemm{
E = {2\pi\over \kappa^2}\left(B_0\partial_r C_0 - B\partial_r C\right).
}
We implicitly assume the limit $r \to r^\star$ exists. The (quasi-local) mass is
\eqn\masrefw{\eqalign{
m & =  AE - Z j,\cr
&  = {\pi \over \kappa^2}\left[2A\left(B_0\partial_r C_0 - B\partial_r C\right) - {ZBC^3\partial_r Z\over A}\right].
}
}

The angular momentum and (analogous or physical) ADM mass are defined by taking $r$ to infinity or sending $B_t$ to infinity \refs{\Abbott, \HawkingA}
\eqn\masadJ{\eqalign{
J = \lim_{r\to\infty}j(r),\cr
M = \lim_{r\to\infty}m(r).
}
}
In section four of the paper we set $\kappa^2 = \pi$ (in some length unit) for simplicity. This is equivalent to $8G_N = 1$ (in some length unit).

\appendix{D}{Kruskal coordinates}

The metrics which we study in this paper are in general of the form $ds^2 = \Delta^2 d{\hat s}^2$, where the factor $\Delta^2$ is (manifestly) positive and regular everywhere.\foot{Except at a curvature singularity.} The line element $d{\hat s}^2$ is of two types. One type has the form (see, \eg, \mmm\ (with \bbbb) and \vvvv) 
\eqn\genmetric{d{\hat s}^2 = \Delta^{-2} f^{-1}d\rho^2 - f dt^2 + \rho^2(d\phi + g dt)^2,
}
and the other has (in general with different $f, g$ and $\Delta$) the form (see, \eg, \newFFF, \spMwww\ and \bux)
\eqn\genmetricP{d{\hat s}^2 = \Delta^{-2} f^{-1}d\rho^2 - f  dt^2 + (d\phi + g dt)^2.
}
The functions $f, g$ and $\Delta$ depend only on $\rho$. $\phi$ is an angular coordinate. The function $f$ has either one or two simple zeros or poles. The scalar curvature is finite at the zero(s). At large $\rho$ both are conformally flat. Thus, each of their Penrose diagrams is similar at large $\rho$ to that of a flat spacetime.

 Around each zero we can introduce a coordinate patch to analytically continue the metric. We now show how to continue the metric \genmetric\ across the zero(s). The analysis is identical for \genmetricP. We follow \Graves\ and \Banadosgq.

We show that around a zero of $f$ which we denote by $\rho_0$ the metric \genmetric\ can be put into the form 
\eqn\genmetricK{d{\hat s}^2 = \Omega^{2}\left(du^2 - dv^2\right) + \rho^2(d\varphi + (g - g_0) dt)^2, \quad \varphi = \phi + g_0t,
}
where $t = t(u,v)$ and $g_0 = g(\rho_0)$. The factor $\Omega^2(u,v)$ is regular at the zero. Thus, \genmetricK\ is a continuation of \genmetric\ across the surface $\rho = \rho_0$. In terms of $u(\rho, t)$ and $v(\rho, t)$ light cones at the zero are lines with slope $\pm 1$. We show \genmetricK\ by finding $u(t, \rho)$ and $v(t, \rho)$.

By comparing \genmetric\ and \genmetricK\ we find 
\eqn\compOT{\matrix{ \Omega^2(u_\rho^2 - v_\rho^2) & = & \Delta^{-2} f^{-1}, \cr
\Omega^2(u_t^2 - v_t^2) & = & -f,\cr
u_\rho u_t - v_\rho v_t & = & 0,
}
}
where $u_\rho = \partial u/\partial \rho$, $u_t = \partial u/\partial t$. Similar notations apply for $v$. We eliminate $\Omega$ and find a relation for $u_t(v_t)$ and $v_\rho(u_\rho)$. We first take a ratio of the first two equations, and then use the last equation to simplify it. The result is
\eqn\equrut{
{u_t^2 - v_t^2\over u_\rho^2 - v_\rho^2} = -{u_t^2\over v_\rho^2} = - f^2\Delta^{2}.
}
Thus,
\eqn\relurvt{u_t = {f \Delta}v_\rho, \quad v_t = {f \Delta}u_\rho.
}
We now introduce a new radial coordinate $\rho^{\star}$, defined by
\eqn\newrst{d\rho^{\star} = f^{-1}\Delta^{-1} d\rho.
}
In terms of $t$ and (the tortoise coordinate) $\rho^{\star}$, the equations \relurvt\ become
\eqn\relurvtP{u_t = v_{\rho^{\star}}, \quad v_t = u_{\rho^{\star}}.
}
The general solution of these equations is 
\eqn\gensoll{u = h_+(\rho^{\star} + t) + h_-(\rho^{\star} - t), \quad v = h_+(\rho^{\star} + t) - h_-(\rho^{\star} - t),
}
where $h_\pm$ are arbitrary functions of (the Eddington-Finkelstein coordinates) $\rho^{\star} \pm t$. We take the Kruskal ansatz,
\eqn\ansatzKU{h_\pm = A_\pm e^{\kappa(\rho^{\star} \pm t)},
}
where $A_\pm$ and $\kappa$ are constants. This gives the factor 
\eqn\factom{\Omega^2 = -{f\over u_t^2 - v_t^2} = {f\over 4A_+ A_- \kappa^2}e^{-2\kappa\rho^{\star}}.
 }
 $\kappa$ is chosen such that $\Omega^2$ is regular, \ie\ non-singular, at the zero of $f$ and on the patch around it. Since the zero or pole is simple, the constant $\kappa$ exists. The Kruskal coordinates are now given by (with $A = A_- = A_+$) 
 \eqn\kursCoo{u = 2Ae^{\kappa \rho^{\star}}\cosh(\kappa t), \quad v = 2Ae^{\kappa\rho^{\star}}\sinh(\kappa t).
 }
 Without loss of generality we can take $A = \pm 1$. The inverse transformations are
 \eqn\invrhostT{\rho^{\star} = {1\over 2\kappa}\ln\left({u^2 - v^2\over 4A^2}\right), \quad t = {1\over 2\kappa}\tanh^{-1}\left({2uv\over u^2 + v^2}\right).
 }
 
 Penrose coordinates $(U, V)$ are given by the usual relations 
 \eqn\penRCK{u + v = \tan\left({U + V\over 2}\right), \quad u - v = \tan\left({U - V\over 2}\right).
 }
 
In a situation where there are more then one zero, the analysis involves introducing several patches, one for each zero. Overlapping regions are identified. The maximal extension requires including and gluing together infinite copies of the patches.
 
 We now give an example to illustrate the procedure. We consider the case \bbbb. We find it convenient to define 
 \eqn\varSI{\lambda = {\rho_-\over \rho_+}, \quad c_1^2 = {l^2\over \rho_+^2(1 + \mu)}\sqrt{{\mu^2 - \lambda^2\over 1 - \lambda^2}}, \quad c_2^2 = {2(1 - \mu^2)\over \mu(\mu^2 - \lambda^2)}, \quad x = {\rho\over \rho_+}, \quad x^{\star} = {\rho^{\star}\over \rho_+},
 }
 where $\mu = (1 + \gamma)/(1 - \gamma)$ (see \bbbb). Note $0 < \lambda < \mu < 1$. In our notation 
 \eqn\variQUQUQ{f = {\rho_+^2\over l^2}\cdot {(x^2 - 1)(x^2 - \lambda^2)\over x^2}, \quad \Delta^{-1} = {\rho_+^2\over l^2}c_1^2\sqrt{1 + c_2^2 x^2}, \quad g = -{\lambda\over l x^2}.
 }
 
  The solution to the equation \newrst\ (for all $x$) is
  \eqn\solnewrst{\matrix{x^{\star} & = & -{c_1^2\over 2(1 - \lambda^2)}\pmatrix{-\sqrt{1 + c^2_2}\ln(|1 - x^2|) +\lambda\sqrt{1 + \lambda^2 c_2^2}\ln(|\lambda^2 - x^2|) \cr
 - 2 |c_2|(1 - \lambda^2)\ln\left(1 + 2c_2^2 x^2 + 2|c_2| x\sqrt{1 + c_2^2 x^2}\right)\cr
 + \sqrt{1 + c_2^2}\ln\left(1 + (1 + 2c_2^2)x^2 + 2\sqrt{1 + c_2^2}\sqrt{1 + c_2^2 x^2}x\right)  \cr
- \lambda\sqrt{1 + \lambda^2 c_2^2}\ln\left(\lambda^2 + (1 + 2\lambda^2 c_2^2)x^2 + 2\lambda\sqrt{1 + \lambda^2 c_2^2}\sqrt{1 + c_2^2x^2}x\right)
 },\cr
& = &|c_2|c_1^2\ln\left(1 + 2c_2^2 x^2 + 2|c_2| x\sqrt{1 + c_2^2 x^2}\right)\cr
& & + {1\over 2\rho_+}\pmatrix{{1\over \kappa_1}\ln(|1 - x^2|) + {1\over \kappa_2}\ln(|\lambda^2 - x^2|) \cr
 - {1\over \kappa_1}\ln\left(1 + (1 + 2c_2^2)x^2 + 2\sqrt{1 + c_2^2}\sqrt{1 + c_2^2 x^2}x\right)  \cr
- {1\over \kappa_2}\ln\left(\lambda^2 + (1 + 2\lambda^2 c_2^2)x^2 + 2\lambda\sqrt{1 + \lambda^2 c_2^2}\sqrt{1 + c_2^2x^2}x\right)
 },
 }
 }
 where $\kappa_1$ and $\kappa_2$ are given by 
  \eqn\kappaaaPU{
\kappa_1 = {1\over \rho_+}\cdot {1 - \lambda^2\over c_1^2\sqrt{1 + c_2^2}}, \quad  \kappa_2 = -{1\over \rho_+}\cdot {1 - \lambda^2\over \lambda c_1^2\sqrt{1 + \lambda^2 c_2^2}}, \quad {\kappa_2\over \kappa_1} < - 1.
 }
Without loss of generality we can take $c_2$ to be positive.

There are two zeros. In the patch around $x = 1$ we must take $\kappa = \kappa_1$ so that the factor $\Omega^2$ is regular at and around $x = 1$. We denote this patch by $K_1$. It covers the region $x > \lambda$. Note on $K_1$ the constant $g_0 = -\lambda/l$. In the patch around $x = \lambda$ however we must take $\kappa = \kappa_2$. This ensures that $\Omega^2$ is non-singular at $x = \lambda$. We denote the patch that covers the region $0 < x < 1$ by $K_2$. On $K_2$ the constant $g_0 = -1/(l\lambda)$. In the overlapping region $\lambda < x < 1$ we identify $K_1$ and $K_2$. We denote the overlap by $K_0$.
 
 In the large $x$ limit we get $x^{\star} = 2|c_2|c_1^2\ln x + h(c_1, c_2, \lambda)$.\foot{The  BTZ solution is obtained by setting $\mu = 1$, \ie, $c_2 = 0$. For $c_2 = 0$, in the large $x$ limit, we instead obtain $x^{\star} = -c_1^2/x + {\cal O}(x^{-3})$.} The function $h$ in particular satisfies $h(c_1, 0, \lambda) = 0$. In the large $x$ limit and at constant $\varphi$ the metric \genmetricK\ up to an overall positive factor equals $du^2 - dv^2 + [\lambda^2/(u^2 - v^2)](udv - vdu)^2$. Therefore, outgoing and ingoing radial null lines end on future and past null infinities, respectively.

Penrose diagrams for the regions $ {\rm III} = \{(x, t)|0 < x \leq \lambda, t\in \IR\}$, ${\rm II} = \{(x, t)|\lambda < x < 1, t \in \IR\}$ and ${\rm I} = \{(x, t)|1 \leq x < \infty, t\in \IR\}$ are given in Fig. 7. The Penrose diagram for the domain III is obtained using $K_2$. Similarly, the Penrose diagram for I is obtained using $K_1$. The Penrose diagram for II can be obtained using $K_0$ or either $K_1$ or $K_2$. Here the Penrose diagram for II is obtained using $K_2$. The Penrose diagram of the whole spacetime is obtained by appropriately gluing together these diagrams. The resulting diagram is given in Fig. 1 in section two.
 
 \ifig\loc{The plots depict the Penrose diagrams of the regions I, II and III. Each point represents a circle or the orbit of the isometry $\partial_\varphi$. Straight lines at $\pi/4$ radians represent null lines. We used the values $\lambda = 0.4, \mu = 0.5, A = 1$, $\rho_+ = 1$ (in string units), $l = 1$ (in string units). For $A = -1$ we simply flips $U \to - U$. Plot a) is the Penrose diagram for III. $x = 0$ is mapped to the vertical line $U = \pi/2$. Plot b) is the Penrose diagram for II. It is obtained using $K_2$. The lines $u = \pm v$ at $x = \lambda$ are mapped to the lines $U = \pm V$. $x = 1$ is mapped to $U \pm V =  \pi$. Plot c) is the Penrose diagram for I. Null lines at $x = 1$ are mapped to $U = \pm V$. $x = \infty$ is mapped to $U \pm V = \pi$.}
{\epsfxsize5.5in\epsfbox{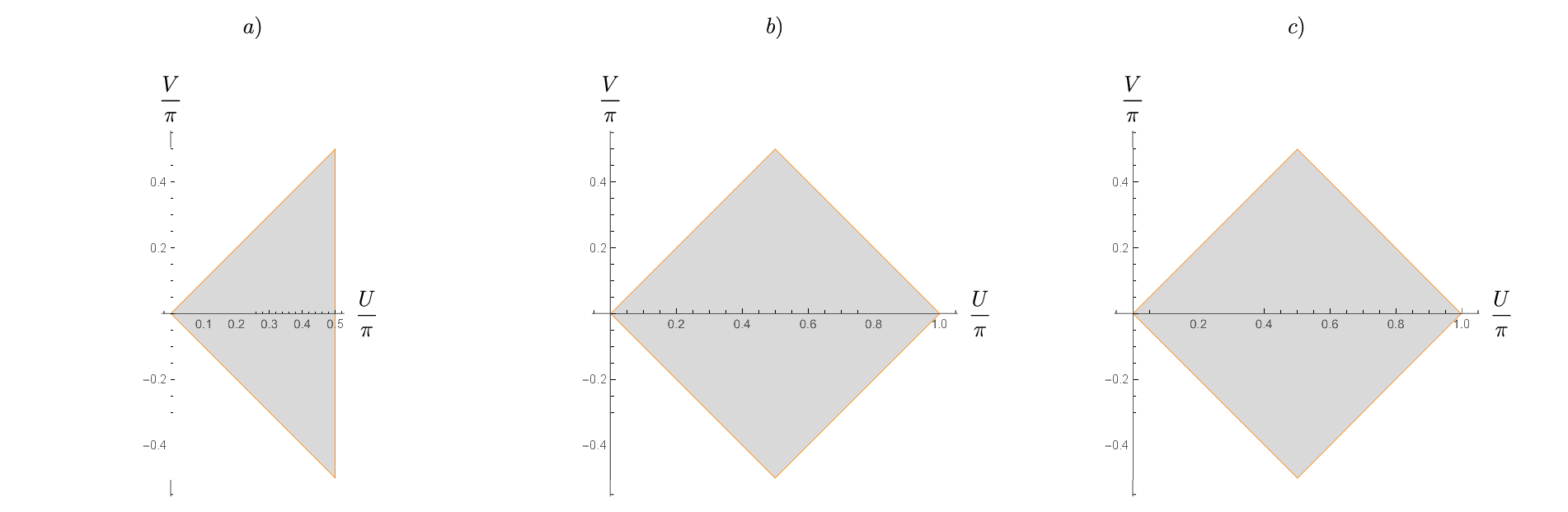}}

\listrefs

\bye